\documentclass[pre,floatfix,twocolumn,10pt]{revtex4-1} 
\usepackage{graphicx}
\usepackage{epstopdf}
\usepackage{natbib}
\usepackage{dcolumn}
\usepackage{bm}
\usepackage{color}
\usepackage[colorlinks=true, linkcolor=blue, citecolor=blue]{hyperref}
\usepackage{subfigure}
\usepackage{psfrag,xspace}
\usepackage{amssymb,amsfonts,amsmath}
\usepackage{setspace}

\begin{document}

\author{A. B. Kolton and E. A. Jagla} 
\affiliation{Comisi\'on Nacional de Energ\'{\i}a At\'omica, Instituto Balseiro (UNCu), and CONICET\\
Centro At\'omico Bariloche, (8400) Bariloche, Argentina}

\title{On the critical region of long-range depinning transitions}

\begin{abstract} 
The depinning transition of elastic interfaces 
with an elastic interaction 
kernel decaying as $1/r^{d+\sigma}$ is characterized by 
critical exponents which continuously vary with $\sigma$. 
These exponents are expected to be unique and universal, 
except in the fully coupled ($-d<\sigma\le 0$) limit, where they depend on 
the ``smooth'' or ``cuspy'' nature of the microscopic pinning potential. 
By accurately comparing the depinning transition 
for cuspy and smooth potentials in a specially devised depinning model,
we explain such peculiar limit in terms of 
the vanishing of the critical region for smooth potentials, 
as we decrease $\sigma$ from the short-range ($\sigma \geq 2$) 
to the fully coupled case. Our results have practical implications 
for the determination of critical depinning 
exponents and identification of depinning universality classes in 
concrete experimental depinning systems with non-local elasticity, 
such as contact lines of liquids  and fractures.
\end{abstract}

\maketitle

\section{Introduction}

Many dissipative disordered systems display a collective 
depinning transition, from an almost static (or inactive) to a sliding (or active) 
regime at a threshold value of a driving force. 
Examples range from field driven domain walls in ferromagnetic 
~\cite{Durin2006,Ferre2013,Durin2016} 
or ferroelectric
materials \cite{Kleemann2007,Paruch2013}, 
crack propagation under stress in heterogeneous materials
\cite{Bonamy2008,Ponson2009}, 
contact lines of liquids on a rough substrate~\cite{moulinet2004,ledoussal2009}, 
imbibition of fluids in porous and fractured media \cite{Planet2009}, 
reaction fronts in porous media~\cite{Atis2015},
solid-solid friction \cite{Bayart2015}, sheared 
amorphous solids or yield stress fluids~\cite{Nicolas2017}, 
dislocation arrays in 
sheared crystals~\cite{sethna2017}, current driven 
vortex lattices in superconductors~\cite{nattermann2000,giamarchi2002,ledoussal2010}, 
skyrmion lattices in ferromagnets~\cite{Schulz2012}, 
to even collective cell migration during wound healing or 
cancer invasion~\cite{Chepizhko2016}. 
The collective nature of the depinning transition 
is often spectacularly manifested at low temperatures 
through some kind of ``crackling noise'' which, well 
beyond the lab scale, much resembles earthquakes, motivating 
also their study within the very same framework~\cite{jagla2010,jagla2014}.

A very fruitful analogy of this problem with equilibrium phase transitions
emerges when
considering the driving force as the control parameter and 
the mean sliding velocity as the order parameter.
This analogy has been useful in
pointing out directions for seeking universal behaviour, 
and inspiring new methodologies~\cite{Fisher1998,Kardar1998}. 
The analogy has also been useful to point out the relevance of
genuine non equilibrium effects~\cite{Amaral1994}, and to detect non standard 
features of the transition~\cite{kolton2006,kolton2009,purrello2017}.
Among the very different models that can be proposed, 
the depinning transition of elastic manifolds in random media has become a 
paradigmatic basic problem as it presents the essential ingredients 
for a non trivial universal behaviour, together with an advantageous 
combination of analytical~\cite{wiese2006} and numerical~\cite{Ferrero2013} tractability. 
Moreover, it is directly relevant for predicting universality classes 
of various concrete systems where the elastic approximation 
can be justified, notably magnetic domain walls and 
contact lines of liquids menisci.

The depinning transition at zero temperature 
of an overdamped elastic interface 
in a random potential
is continuous, non hysteretic, and occurs at a 
characteristic threshold force $f_c$. 
Close enough and above the threshold 
the mean velocity $v$ of the interface in the direction 
of the force is well 
described by the putative depinning law 
$v\sim (f-f_c)^{\beta}$, with $\beta$ a non-trivial 
critical exponent. A divergent correlation 
length $l \sim (f-f_c)^{-\nu}$ and a divergent 
correlation time $\tau \sim l^{z}$ characterize the 
jerky motion as we approach $f_c$ from above. 
Concomitantly, the rough geometry of the interface 
becomes self-affine with the displacement field growing 
as $u \sim x^{\zeta}$ for length-scales $x$ below $l$. 
Hence $v\sim l^{\zeta-z}$ and $\beta=\nu(z-\zeta)$. 
In this regime, the spatio-temporal fluctuations 
also display universal behaviour and are controlled by avalanches 
with a broad distribution of sizes $S$ 
(and durations $T \sim S^{z/(1+\zeta)}$), 
such that $P(S)\sim S^{-\tau}$, with $\tau=2-(\zeta+1/\nu)/(d+\zeta)$
in the quasistatic limit. 
The critical exponents can be estimated 
analytically \cite{ledoussal2002,Fedorenko2003} and numerically 
\cite{Leschhorn1993,Leschhorn1997,Roters1999,Rosso2003,Rosso2007,Ferrero2013b} 
to determine the different universality classes. 
These are determined by $d$, the 
range~\cite{Ramanathan1998,Zapperi1998,Rosso2002,Duemmer2007,Laurson2013} 
or nature~\cite{Boltz2014}
of the elastic interactions, the anisotropic~\cite{Tang1995} 
or isotropic correlations of the 
pinning force~\cite{Fedorenko2006,Bustingorry2010}, 
and by the presence of additional 
(i.e. apart from the pinning force) non-linear 
terms~\cite{Amaral1994,Tang1995,Rosso2001b,Goodman2004,ledoussal2003,Chen2015}. 
Boundary ~\cite{Aragon2016} or ac driven \cite{glatz2003} 
depinning of elastic interfaces have been also studied.
If the so called statistical tilt symmetry holds, 
only two exponents are needed to fully characterize 
the depinning universality class. 
In any case, it is very convenient to consider separately the 
purely geometric $\zeta$ or $\nu$, which do not 
involve time scaling, from $z$ or $\beta$ which do.

In order to quantify the universal properties of the depinning transition 
for a concrete experimental system (or microscopic model with a 
yet unknown coarse grained dynamics) it is important to 
determine the critical exponents accurately enough so 
to be able, at least, to differentiate between 
different candidate universality classes. 
Unfortunately, 
testing the depinning law is in general a rather difficult 
task experimentally, and in many cases also numerically.
On one hand~\footnote{We already discard the possible existence of 
finite-size ~\cite{Duemmer2005} and other type of 
crossovers (see e.g. Refs.~\cite{Bustingorry2010,Bustingorry2010b,Chen2015}), 
or thermal rounding effects~\cite{Chen1995,Bustingorry2012,purrello2017} 
which may also difficult the experimental observation 
of a clear power-law in the velocity force characteristics.},
fitting accurately $\beta$ certainly requires an accurate estimation 
of the non-universal threshold force $f_c$. 
In that respect it is important to note that 
the depinning force 
$f_c$ displays important sample to sample 
fluctuations in finite systems~\cite{Bolech2004}, 
with $\overline{[f_c-\overline{f_c}]^2} \sim L_0^{-2/\nu_{FS}}$, 
where $L_0$ is the linear size of the system and $\nu_{FS} \geq 2/(d+\zeta)$~\cite{ledoussal2002}. The 
thermodynamic 
limit $L_0\to \infty$ is also delicate as the value of $f_c$ can be 
strongly affected by 
the anisotropic sample aspect-ratio scaling we keep 
in such limit~\cite{Bolech2004,Kolton2013}. 
On the other hand, 
even if we are able to get a sharply defined $f_c$, we are faced to 
the fact that the depinning law is expected to work only in an 
unknown critical region of size $\sim \Delta f^{\tt crit}$, such that the
asymptotic power 
law scaling for $v$ fully develops 
only for $(f-f_c) \lesssim  \Delta f^{\tt crit}$.  
Knowing roughly $\Delta f^{\tt crit}$ is thus 
fundamental for practical applications of the theory.
Little is known however about $\Delta f^{\tt crit}$ for the depinning transition, 
except that it is non-universal. 
How does the critical region depend on the microscopic
shape of the disorder, the range of the elastic interactions or the 
dimensionality $d$? Do scaling corrections produce intermediate 
power-laws with effective exponents? If so, are the effective 
exponents expected to be larger or smaller than the true ones?
Do they violate the expected asymptotic scaling relations 
among exponents?

In this paper we try to answer some of the above mentioned 
practical questions 
by performing numerical simulations on different 
microscopic models. We study  
depinning models with  isotropic 
uncorrelated disorder and harmonic long-range 
elasticity, with elasticity kernel decaying with distance as 
$1/r^{d+\sigma}$. We vary the range 
of the elastic interactions from the ($\sigma\le 0$) fully 
coupled case to the ($\sigma \geq  2$) short-range cases 
and compare the critical behaviour of the 
velocity-force characteristics 
for two different forms of the  microscopic disorder. 
They are termed the ``smooth'' case (in which the force originated in the disorder 
does not have any discontinuities), and the "cuspy" case (in which the force has 
an abrupt jump at the transition point between different potential basins).
For the cuspy potential the extent of the critical region tends to be large and rather independent on the value of $\sigma$.
For the smooth potential  the 
critical region where universality holds (i.e. where we get the same exponents 
than in the the cuspy case), decreases by increasing 
the range of elastic interactions 
and strictly vanishes, $\Delta f ^{\tt crit} \to 0$, in the fully coupled limit.
In such limit scaling corrections are not anymore ``corrections'' but control 
the ultimate asymptotic scaling. This
explains the peculiar non-universality of the 
fully coupled model in the strong pinning phase (i.e. with $f_c>0$),
which displays two different exponents, $\beta=3/2$ and $\beta=1$, 
for the smooth and cuspy cases, respectively (as it is well 
known from the equivalent and exactly solvable Prandtl-Tomlinson model). 
For $\sigma>0$, where a unique value of $\beta$ is the ``right'' one
\footnote{It is interesting in this respect to quote the following statements by Fisher, in 
Ref.\onlinecite{Fisher1998}: ``The velocity exponent, $\beta$, in mean field theory is not fully universal. It depends on whether or not the pinning forces are continuously differentiable and, if not, on the nature of the singularities in
$f_p(u)$. For all but discontinuous forces, this
causes long time scales in mean-field theory associated with the acceleration away from configurations that have just
become unstable. With short range interactions, these should not play a role due to the jerky nature of
$\phi({\bf r},t)$ caused by jumps. Thus the mean-field models with $\beta_{MF}=1$ are the \textit{right} ones about which to attempt an RG analysis.''
}
, 
our results show nevertheless the great importance of scaling 
corrections and the emergence of dangerous effective power-laws 
which particularly affect the obtention of the 
asymptotic dynamical exponents $\beta$ or $z$ as compared with 
the roughness exponent $\zeta$ which is found to be more robust.
These corrections are particularly relevant for a successful experimental 
(and also numerical) identification of depinning universality classes 
in elastic systems with long-range interactions ($0<\sigma<2$) 
and to explain quantitative discrepancies with theory. 
To arrive to these results we devise a convenient 
model for comparing the critical behaviour of 
cuspy and smooth microscopic disorders accurately.

\section{Generalities of the basic model}

We model a $d$-dimensional interface embedded on a $d+1$ disordered material as a collection of blocks
$i=1,\ldots, N$, 
located at sites of a $d$-dimensional regular lattice, and characterized by a 
continuous displacement $u_1,\ldots,u_{N}$ in the $d+1$ transverse direction.
We will assume an overdamped Equation of motion, 
\begin{equation}\label{eqmotion}
\dot{u}_i(t)=\sum_{j=1}^{N} G_{ij} \, (u_j- u_i) + F_i(u_i) + f.
\end{equation}
where the terms on the right hand side represent the sum of the elastic couplings, the disorder, and the 
uniform and constant pulling force, respectively.

The $G$ term in Eq. (\ref{eqmotion}) accounts for the harmonic elastic interactions, 
with $G_{ij}$ being the spring constant 
associated with the blocks $i$ and $j$. In order to model long-range elastic interactions we 
use $G_{ij} =\kappa/|i-j|^{d+\sigma}$, 
with the normalizing constant $\kappa$ used to obtain $\sum_{j} G_{ij}=1$ (note that the value of $G_ {ii}$
does not influence Eq. (\ref{eqmotion}), and is taken as zero). 
The $G$ term just described implies the convex elastic 
energy $\sum_{ij} G_{ij}(u_i-u_j)^2/2$.

When $\sigma \geq 2$ the elastic kernel represents a short-range elastic interaction, while  
for $-d<\sigma\le0$, it 
represents the fully coupled case that can be exactly solved using mean field techniques. 
Periodic boundary conditions can be taken into account 
by summing the elementary kernel over 
periodic images of the finite system 
and by using its Fourier 
representation to obtain the 
elastic forces at each step of time integration
through a numerically efficient convolution.

\begin{figure}
\includegraphics[width=\columnwidth,clip=true]{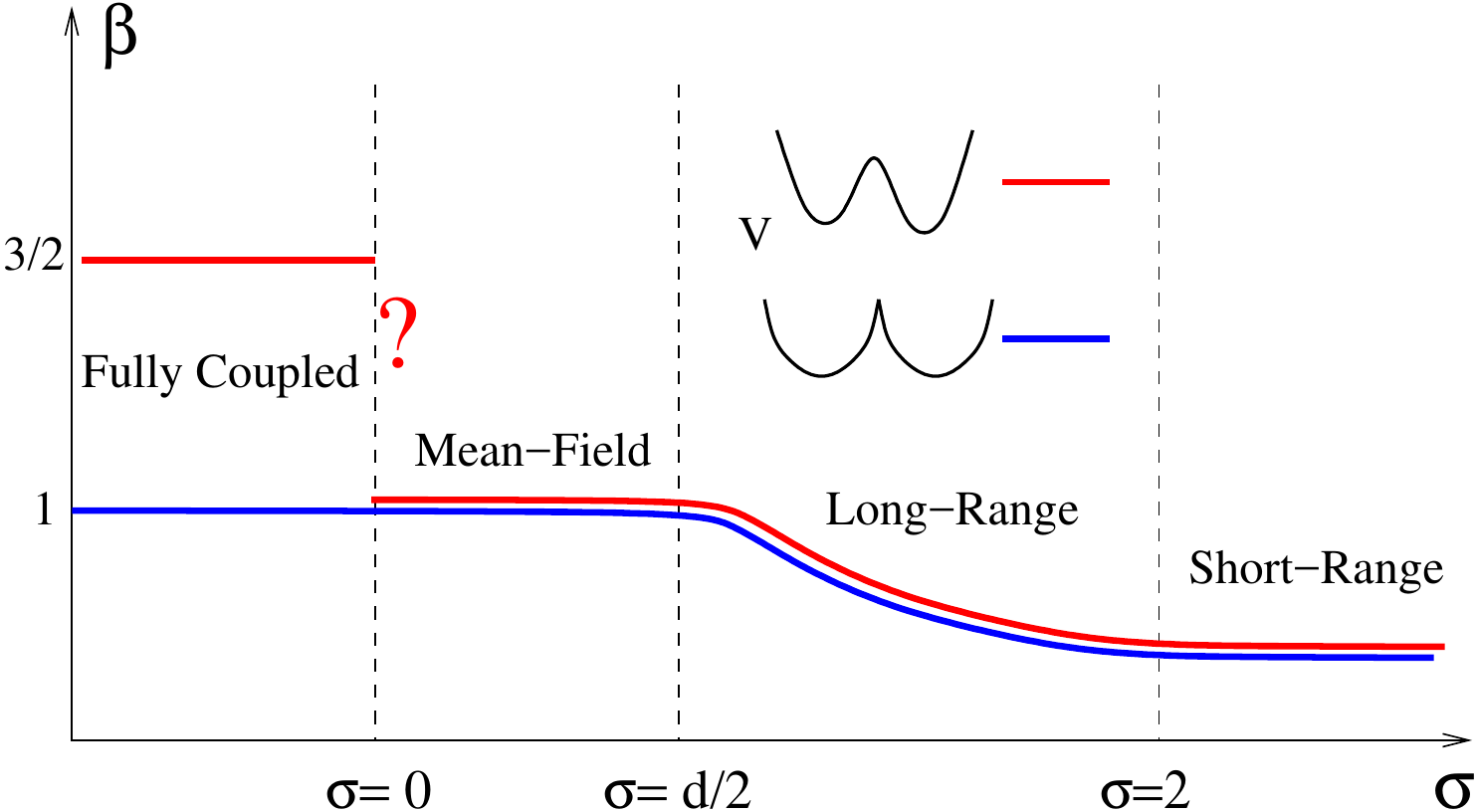}
\caption{Schematic illustration of the peculiar, non-universal, fully-coupled limit 
in the strong pinning phase. 
The velocity exponent $\beta$ is universal (i.e. independent 
of the shape of the microscopic potential), from its 
short-range value for $\sigma \geq 2$, varying continuously 
to its mean-field value for $\sigma \leq d/2$. 
In the region $\sigma \le 0$ the universality is suddenly broken: 
the smooth potential $\beta$ (solid red line) does not coincide 
with its cuspy potential value (solid blue line).
How does this affect the critical region for 
the long-range depinning transitions? 
\label{fig:question}
}
\end{figure}
The second term in Eq.(\ref{eqmotion}) (the only non-linear term of the equation of motion) 
accounts for the pinning forces.
For the moment we will just assume it is statistically characterized by 
$\overline{F_{i}(u)}= 0$, 
$\overline{F_{i}(u) F_j(u')}= \Delta(u-u')\delta_{ij}$, where 
$\overline{ \cdots} $ stands for the average over disorder realizations and $\Delta(u)$ is a short-ranged 
function with $\Delta(0)$ measuring the strength of the disorder.

The motion described by Eq.(\ref{eqmotion}), 
with its convex elastic energy 
is characterized 
by a unique critical force $f_c$ in the large-size limit
~\footnote{We assume that $f_c$ 
is not controlled by extreme statistics in the large-size 
limit~\cite{Kolton2013}.}. 
This critical force is important in determining the 
fate of the system at very long times. 
If $f<f_c$ the system reaches a static solution 
such that $\dot{u}_i(t)=0$ $\forall i$.
For $f>f_c$ it reaches a unique steady-state with
$\dot{u}_i(t)\geq 0$ univocally defined up to a global time shift, 
with the mean velocity defined as 
$v \equiv N^{-1} \overline{\sum_{j=1}^{N} \dot{u}_i(t)}$ or, 
by $v \equiv \lim_{t\to \infty} 
 {(Nt)^{-1} \sum_{j=1}^{N} (u_i(t)-u_i(0))}$ thanks 
to self-averaging.
The properties just described are consequences of 
the Middleton theorems~\cite{Middleton1992} 
which assure that the dynamics described by 
Eq.(\ref{eqmotion}) with its convex elastic energy 
converges for $f \geq f_c$ to a unique ``Middleton attractor''. 
The properties of this attractor 
can be exploited to devise smart algorithms 
to target the critical force and critical 
configuration in finite samples without solving the true dynamics
~\cite{Rosso2001,Rosso2005}.
It also allows to cleanly visualize the convergence towards 
the steady-state by reparametrizing the time 
with the system center of mass 
$u(t) \equiv  N^{-1} \sum_{j=1}^{L^d} u_i(t)$~\cite{Kolton2009b}.
In this paper we will 
rely (apart from 
the unicity of the dynamical attractor) on the 
general property $\dot{u}_i(t)\geq 0$, valid 
in the $f \geq f_c$ steady-state.
This will be particularly important in relation to 
the model discussed in Sec.\ref{sec:discrete}.

It is also worth remarking here that the so called 
statistical tilt symmetry (STS) holds for 
Eq.(\ref{eqmotion}), so $\nu = 1/(\sigma-\zeta)$ for $d/2 \leq \sigma \leq 2$, 
$\nu = 1/(2-\zeta)$ for $\sigma \geq 2$, and $\nu = 1/2$ for $\sigma < d/2$~\cite{Kardar1998}. 
Therefore only two exponents are needed to fully characterize 
the depinning universality class. 
\footnote{The discussion of anisotropic depinning universality classes, 
where STS is broken will be published elsewhere. 
Nevertheless, we believe that our general conclusions 
hold also for this case.}
Using the STS, in this work we will consider  
separately $\zeta$ and $\beta$, to characterize the universality classes. 
As we will discuss later this 
arbitrary separation is nevertheless quite convenient, as $\zeta$ (and $\nu$) 
has a purely geometric origin unlike $\beta$ (and $z$) 
which are related with time-scaling and thus affected by local 
non-universal bottlenecks. 
This has important consequences from the practical point of view.
For instance, it is easier to get much more accurate values for 
$\zeta$ (or $\nu$) by exploiting different methods which do not involve 
a true temporal evolution, such as the variant Monte Carlo 
algorithm~\cite{Rosso2001,Rosso2005} or the metastable configurations obtained 
by relaxing a flat configuration below the 
depinning threshold~\cite{kolton2006b}.

For $f \gg f_c$ the effect of the disorder can be treated as a perturbation. 
At first order disorder mimics an effective temperature proportional 
to $v^{-1}$. In this so called ``fast flow'' 
regime the interface  
can be fairly described by the forced Edwards-Wilkinson equation and 
$v \approx f$~\cite{chauve2000}. 
For the low velocity critical regime $f \gtrsim f_c$ we are interested 
in, perturbation theory fails.
Numerical simulations and the functional renormalization group (FRG) 
approach applied to Equation ~(\ref{eqmotion})  
teach us that, for $\sigma>0$, the above 
description uniquely determines the critical depinning exponents of the model. 
Their values depend on $d$, and smoothly evolve with decreasing 
$\sigma$, from their short-range values for $\sigma \geq 2$, 
to the mean field value for $\sigma \leq d/2$
or equivalently $d \geq d_c(\sigma)$, 
with $d_c=2\sigma$ the upper 
critical dimension~\footnote{The $\sigma=d/2$ case has $\beta=1$ but with weak
logarithmic corrections ~\cite{Fedorenko2003}.}. 
The exact ``shape'' of 
the microscopic pinning force $F_i$ is believed to be unimportant in many respects. 
Indeed, FRG tells us that for the model of Equation ~(\ref{eqmotion}), the bare correlator 
of the pinning force $\Delta(u)$ flows, under coarse graining above 
the fundamental Larkin scale $L_c$, towards a correlator with a ``cuspy'' singularity. 
The existence of such cusp nicely accounts for the existence of a critical force,
and also for the existence of avalanches. 
The fixed point of the renormalization flow equations for 
the pinning correlator function gives us access to unique values of 
$\beta \equiv \beta(d,\sigma)$ and $\zeta \equiv \zeta(d,\sigma)$, 
which completely characterize the depinning universality class of our model.
These FRG calculations are performed assuming in principle 
a small separation $\epsilon \ll 1$ 
from the upper critical dimension, $\epsilon=d_c(\sigma)-d$, 
with $d_c=2\sigma$. One may thus question their validity for 
the experimentally relevant case $d=1$ for instance.
Numerical simulations by Rosso \textit{et.al.} \cite{Rosso2007}
fairly confirm however the FRG picture for 
one-dimensional interfaces with 
short-range elastic interactions 
($\sigma \geq 2$), showing its validity for the 
extreme $\epsilon=3$ case.

The above picture, valid for $\sigma>0$ and in principle
any dimension $d\geq 1$, sharply contrasts with the $\sigma \le 0$ fully coupled limit, where 
the depinning model becomes equivalent to the exactly solvable one-particle 
Prandtl-Tomlinson model, one of the most popular models in nanotribology~\cite{Popov2014}. 
For the strong pinning phase of this model, which has $f_c>0$ (see e.g. Ref.~\cite{Cao2018}), 
the critical behaviour of the velocity $v\sim (f-f_c)^\beta$ 
becomes non-universal for different microscopic potentials. 
On one hand for a smooth random potential $V_i$ such that 
$F_i(u)=-(d/du)V_i(u)$ does not have jumps, one has $\beta=3/2$. 
On the other hand, for a cuspy random potential with force discontinuities 
$\beta=1$ is obtained, a value that coincides with the mean field value 
expected from FRG for $d \leq d_c(\sigma)$. One may thus ask: 
what is exactly happening in the 
$\sigma=0+$ limit of the smooth potential case? (See Fig.\ref{fig:question})

\section{Organization of the paper}
To clarify the last issue and discuss its practical consequences 
we will focus on the $d=1$ model with long-range interactions 
(the $d>1$ short range case will be discussed in the appendix).
First, in Section~\ref{sec:continous} we will consider the 
standard model with continous displacements, and compare 
the critical behaviour of smooth and cuspy microscopic pinning potentials. 
This will allow us to illustrate the kind of effects that can be 
expected in the critical region for $\sigma>0$ in both cases. 
In Section~\ref{sec:discrete} we will propose an alternative model to compare the 
same two pinning cases but much more accurately, using discrete 
displacements and an effective microscopic 
potential described by traps and suitable transition rates. 
In Section~\ref{sec:conclusions} we summarize our results and 
discuss their practical implications for the study of 
the depinning transition.   

\section{Results}
\label{sec:results}

\begin{figure}
\includegraphics[width=7cm,clip=true]{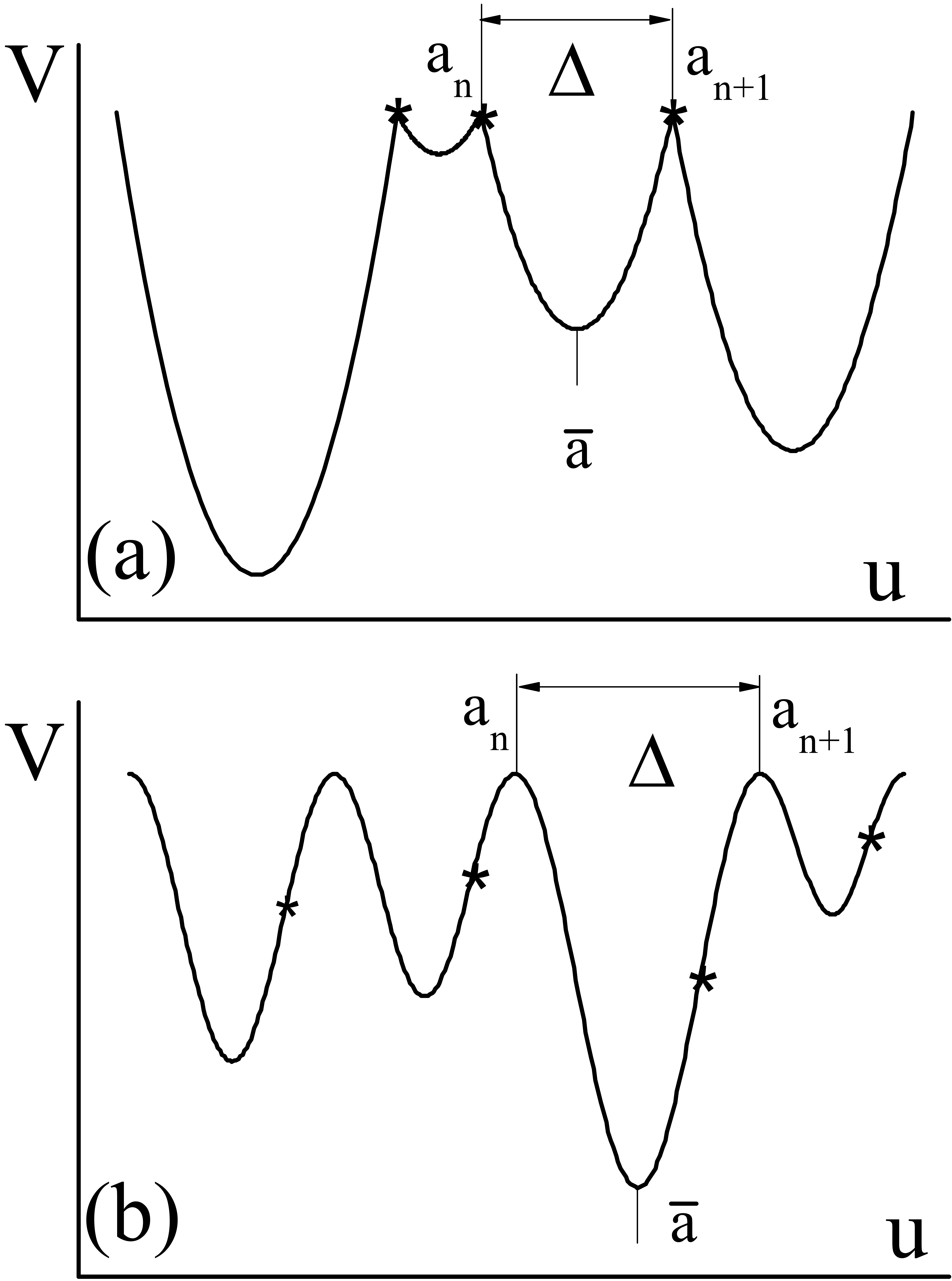}
\caption{Typical forms of the pinning potentials we analyze. Transition points for a particle moving to the right are indicated by stars. (a) A cuspy potential, in which there is a jump of the force at the transition point. (b) A smooth potential, in which the potential itself and its derivatives (up to second order at least) are continuous. 
\label{fig:potenciales}
}
\end{figure}

\begin{figure}
\includegraphics[width=7cm,clip=true]{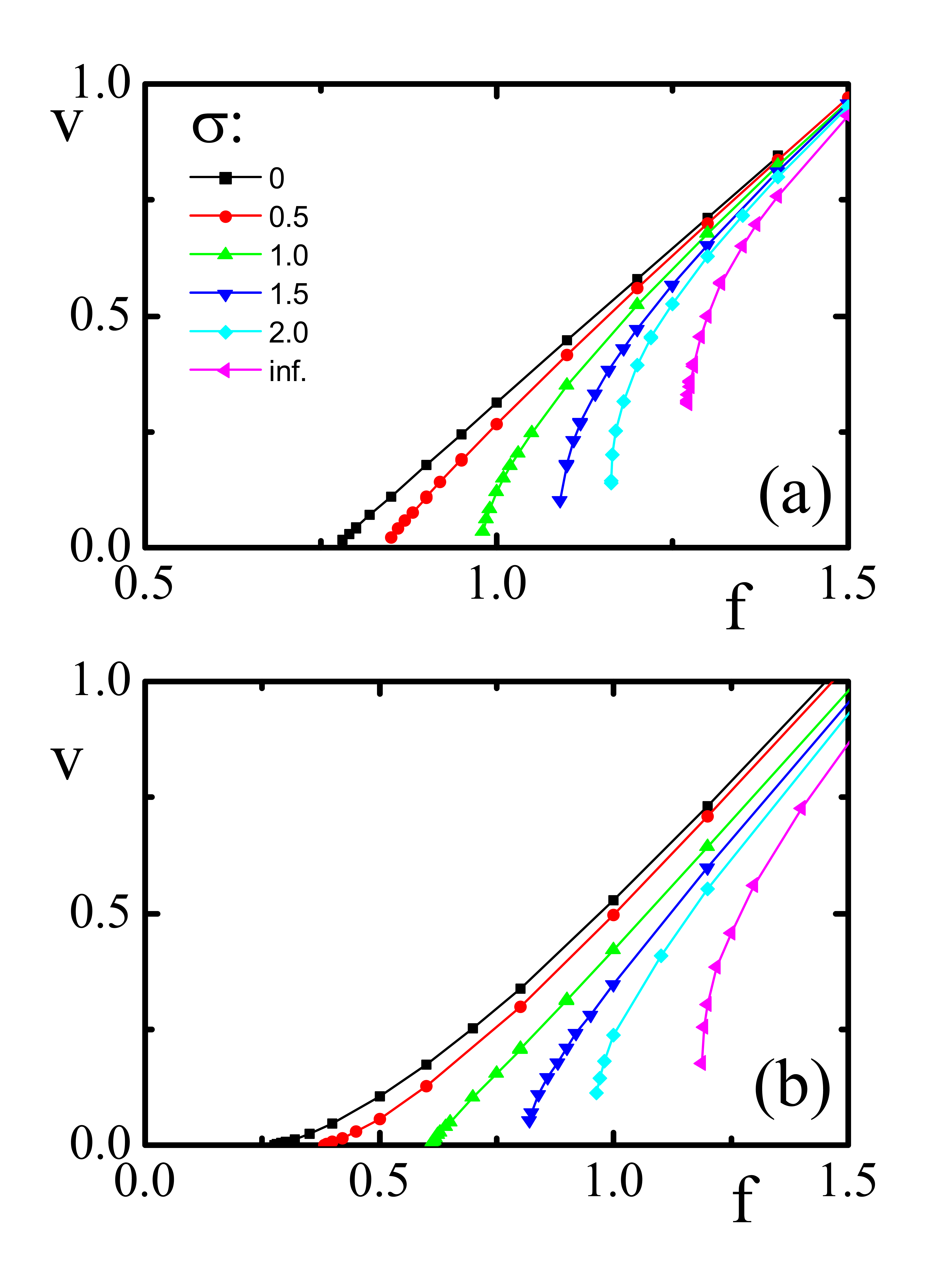}
\caption{Velocity as a function of force for a systems of $N=2^{16}$ sites, for different values of $\sigma$, using cuspy (a) and smooth (b) pinning potentials. Curves with the same value of $\sigma$ display apparently larger values of $\beta$ in (b) than in (a). 
\label{f1}
}
\end{figure}

\subsection{Numerical results: Continuous potentials}
\label{sec:continous}
It is convenient to see the actual results of direct numerical simulations of Eq. \ref{eqmotion} to 
have a first clear picture of the differences that appear between smooth and cuspy pinning potentials.

In concrete, the numerical potentials we used are defined as follows (see Fig. \ref{fig:potenciales})
For each site $i$ a potential $V_i(u_i)$ is constructed. The generic potential $V(u)$ is constructed piecewise, by dividing the $u$ axis in segments through a set of values $a_n$. In each interval $a_n$-$a_{n+1}$ (defining $\overline a\equiv (a_{n+1}+a_{n})/2$, and $\Delta \equiv a_{n+1}-a_{n}$) the potential is defined as
\begin{equation}
V(u)=\left [(u-\overline a)^2-\Delta^2\right ]
\end{equation}
for the cuspy case, and
\begin{equation}
V(u)=-\frac{3\Delta^2}{2\pi^2} \left [1+\cos\left(\frac {2\pi (u-\overline a)}\Delta\right)\right ]
\end{equation}
for the smooth case. Note that even in the smooth case the potential is not analytic, but it has a continuous second derivative, which is enough for our purposes. 
The separation $\Delta$ between $a_n$ and $a_{n+1}$ is stochastically chosen from a flat distribution between $\Delta_{min}=1$ and $\Delta_{max}=2$.

\begin{figure}
\includegraphics[width=7cm,clip=true]{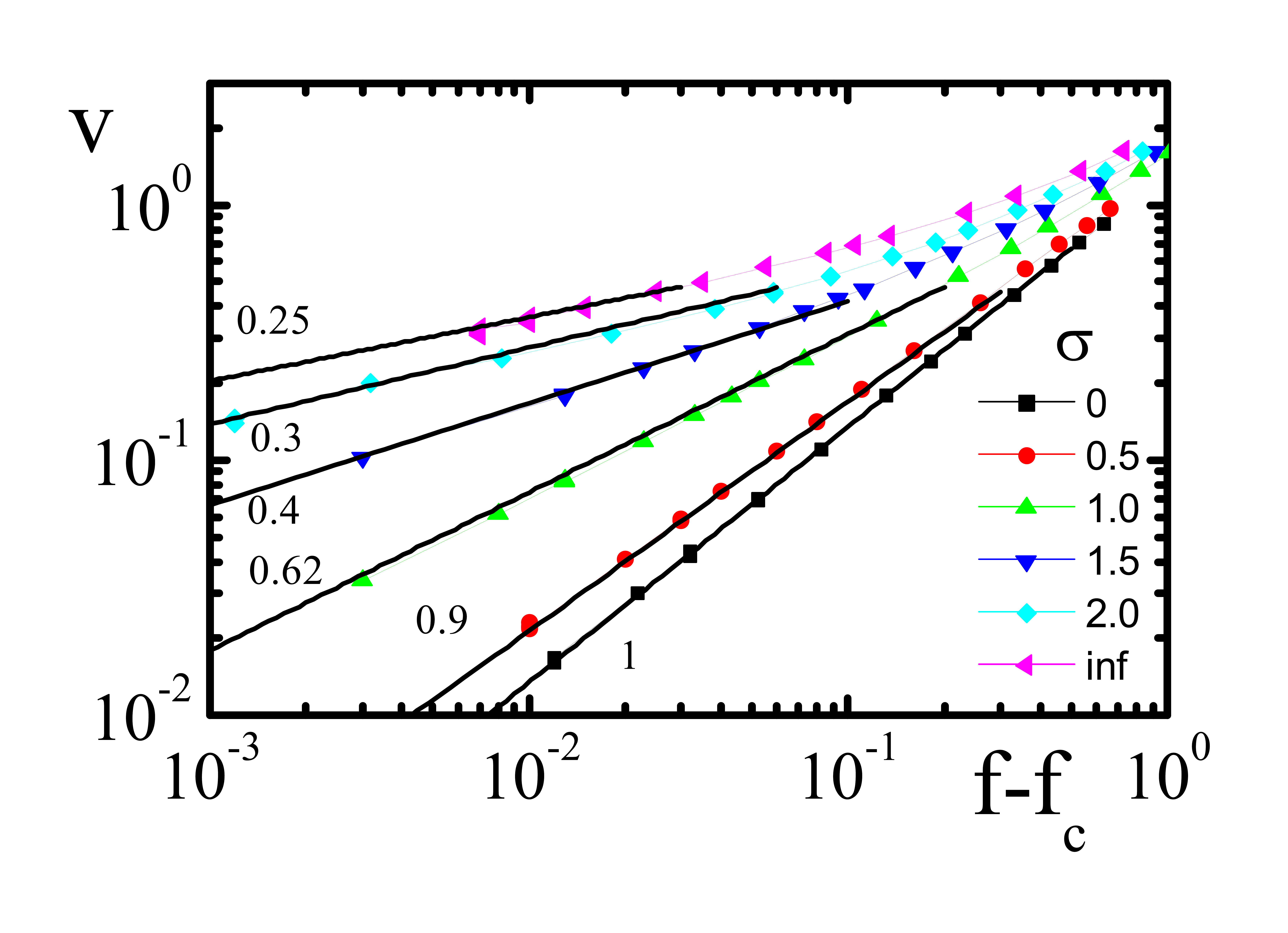}
\caption{
The data in Fig. \ref{f1}(a) plotted in logarithmic scale, fitting the critical force for each value of $\sigma$. The value of the $\beta$ exponent is obtained as the slope of the asymptotic straight behavior, as indicated.
\label{f2}
}
\end{figure}

In Fig. \ref{f1}(a) we see the value of $v$ as a function of $f$ for  the case of cuspy potentials for a few values of $\sigma$ going from nearest neighbor interaction ($\sigma\to\infty$) to mean field interaction ($\sigma\to 0$).
The plot in logarithmic scale with respect to the critical force $f_c$ (Fig. \ref{f2}) (fitting in each case the value of $f_c$) displays a robust critical region in which the $\beta$ exponent can be defined. The value of $\beta$ as a function of $\sigma$ (reported also in Fig. \ref{f2})
increases when $\sigma$ moves from large values to $\sigma=0$. Moreover, the actual values of $\beta$ obtained for different $\sigma$ accurately fit those known from the literature.\footnote{The value for $\sigma=2$, is somewhat lower than the expected $\beta=1$, however this is not surprising in view of the logarithmic corrections expected in this case (see \cite{Fedorenko2003}).}
This represents the ``standard" behavior that is compatible with the analysis using renormalization group techniques.

The results of simulations using smooth potentials are shown in Fig. \ref{f1}(b). They apparently 
show larger values of $\beta$ than the values for cuspy potentials at the same $\sigma$ (Fig. \ref{f1}(a)). For instance, the curve for $\sigma=1$ seems to have a slope close to one, instead of displaying $\beta\simeq 0.62$ as in the cuspy case.

As we have anticipated, the way out of this conundrum is to realize that the critical region at which the results for smooth potentials should coincide with those of cuspy potentials may be small, and we might not be observing it in Fig. \ref{f1}(b). The critical region should become observable when plotting the results of Fig. \ref{f1}(b) in logarithmic scale with respect to the critical force $f_c$. 
However, the identification of the critical region and the true value of $\beta$ relies on the accurate determination of the value of $f_c$, and this has to be done at the same time that fitting the value of $\beta$, so it is very difficult to get reliable values of $\beta$ if the critical region is expected to be very small.
To overcome this difficulty and try to set this point, we have done simulations in a modified model, that is described in the next section.

\subsection{Numerical results: Discrete pinning potential}
\label{sec:discrete}

The results for the flow curves contained in the previous section (Figs. \ref{f1} and \ref{f2}) suggest that there are strong non-universal effects associated to the form of the pinning potential that is used. These non-universal effects can mask the true critical behavior (that must be independent of the form of the potential for $\sigma>0$), and so they have a great practical importance. Yet to determine accurately the behavior close to the critical force $f_c$ we face the problem mentioned at the end of the previous section:
The value of the critical force is not known in advance, and it has to be determined during the fitting process itself. This may be quite inconvenient when the extent of the critical region is very small, as a slight uncertainty in the critical force can completely alter the results in this critical region.

In this Section we present a modification of the model used in the previous Section \ref{sec:continous} that does not have this drawback, and allows a more precise characterization of the effects we are studying. In addition, it also allows a very precise determination of other exponents of the transition, in particular the dynamical exponent $z$, something that we are also interested in.

We start by defining this modified model, that was already introduced in the context of thermal creep 
in ~[\onlinecite{purrello2017}].
Its main characteristic is that the position of the interface at each spatial location is not continuous (as in the previous section) but discrete. 
We first define the applied force $f^{\tt app}$ at site $i$ as
\begin{equation}
f_i^{\tt app}=\sum_{j=1}^{N} G_{ij} \, (u_j- u_i) + f.
\label{fapp}
\end{equation}\
The interface jumps between successive discrete positions (that are taken completely random, with an average separation of 0.1) when $f_i^{\tt app}$ exceeds the local critical force, $f_i^{\tt th}$ (for simplicity in the simulations we take the values of $f_i^{\tt th}$ as constant: $f^{\tt th}=2.5$). 
The jump itself is considered to be instantaneous, but the transition between consecutive positions does not occur immediately after the critical force is exceeded. Instead, a transition rate is considered. In this version of the model, the cuspy and smooth cases of the model in the previous section can be simulated by considering different forms of the transition rate. In fact, as it will be further discussed below, the cuspy and smooth cases of the previous section differ mainly in the time it takes for a particle that has exceeded the stability limit of one potential well to reach the next equilibrium point at the next well. In the case of cuspy potentials this time is roughly constant, independent of the extent by which the threshold force has been exceeded. In the case of smooth potentials this time goes as $\sim (f-f^{\tt th})^{-1/2}$, typical of saddle-node bifurcations. We can model this behavior by assigning a constant transition rate $\lambda\sim \text{cte}$ to mimic the effect of cuspy potentials (this will be referred to as the ``constant rate'' case), and a rate that depends on applied force as  
$\lambda \sim (f-f^{\tt th})^{1/2}$ to 
simulate  the case of smooth potentials (this will be referred to as the ``variable rate'' case). 
In the concrete implementation, we consider all unstable sites for which $f_i^{\tt app}>f_{\tt th}$, and calculate an expected time $\tau_i$ for each site to jump, taken from a Poisson distribution with the corresponding rate $\lambda_i$. The lowest of all $\tau_i$ is chosen and this is the site that is actually moved. Time is advanced by this minimum $\tau_i$, elastic forces are re-calculated and the process is continued. Average velocity is simply calculated as $(u(t)-u(0))/t$, in a run over a long time interval $t$.

The advantage of the discrete potential model is that its quasistatic properties are completely independent of the rate law that is used. In fact, let us consider for instance two avalanches in the system that start from the same configuration  but that evolve according to two different rate laws. Since the effect of any forward jump of any portion of the interface is to increase the force over any other site in the system, the avalanches will be exactly the same whatever the rate law is, the only possible difference is in the order and time of activation of different sites, by virtue of the Middleton theorems. This means at once that all static critical exponents such as $\tau$, $\zeta$, $\nu$... must be independent of the rate law.
In addition, the critical force will also be independent of the rate law, which is very convenient from the point of view of accuracy of the simulations, as explained previously. The only exponents that can depend on the rate law are those that sense temporal properties of the dynamics. They are the flow exponent $\beta$ and the dynamical exponent $z$.

\begin{figure}
\includegraphics[width=7cm,clip=true]{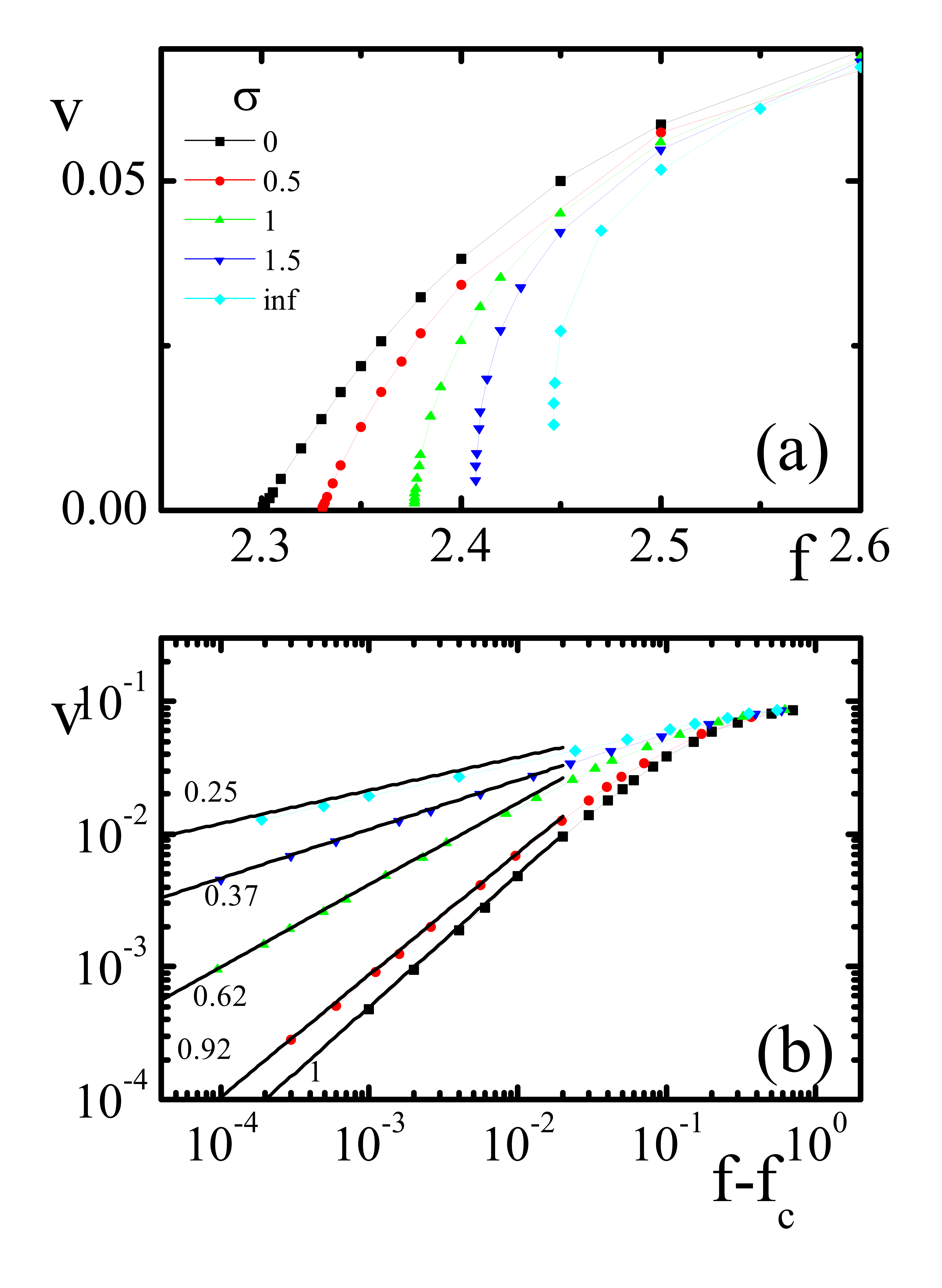}
\caption{(a) Flow curves of the discrete pinning potential model with constant transition rate for different values of $\sigma$. 
(b) Same data in logarithmic scale, fitting the value of $f_c$ in each case. From these curves the value of 
the $\beta$ exponent can be determined.
\label{lambda_cte}
}
\end{figure}
The flow curves obtained at constant rate for different values of $\sigma$ are shown in Fig. \ref{lambda_cte}.
There is a clear difference in the flow curves between Figs. \ref{lambda_cte} and \ref{f1} for large values of $f$, which is a consequence of the details of the models (the discrete pinning model does not have a ``fast flow'' regime, but a velocity saturation at large forces. However, the values of the $\beta$ exponent determined from the logarithmic plots (panel (b)) are in excellent agreement with those in Fig. \ref{f2}, showing that the discrete model with constant rated in fact reproduces the behavior of the continuous cuspy pinning potential. 
\begin{figure}
\includegraphics[width=7cm,clip=true]{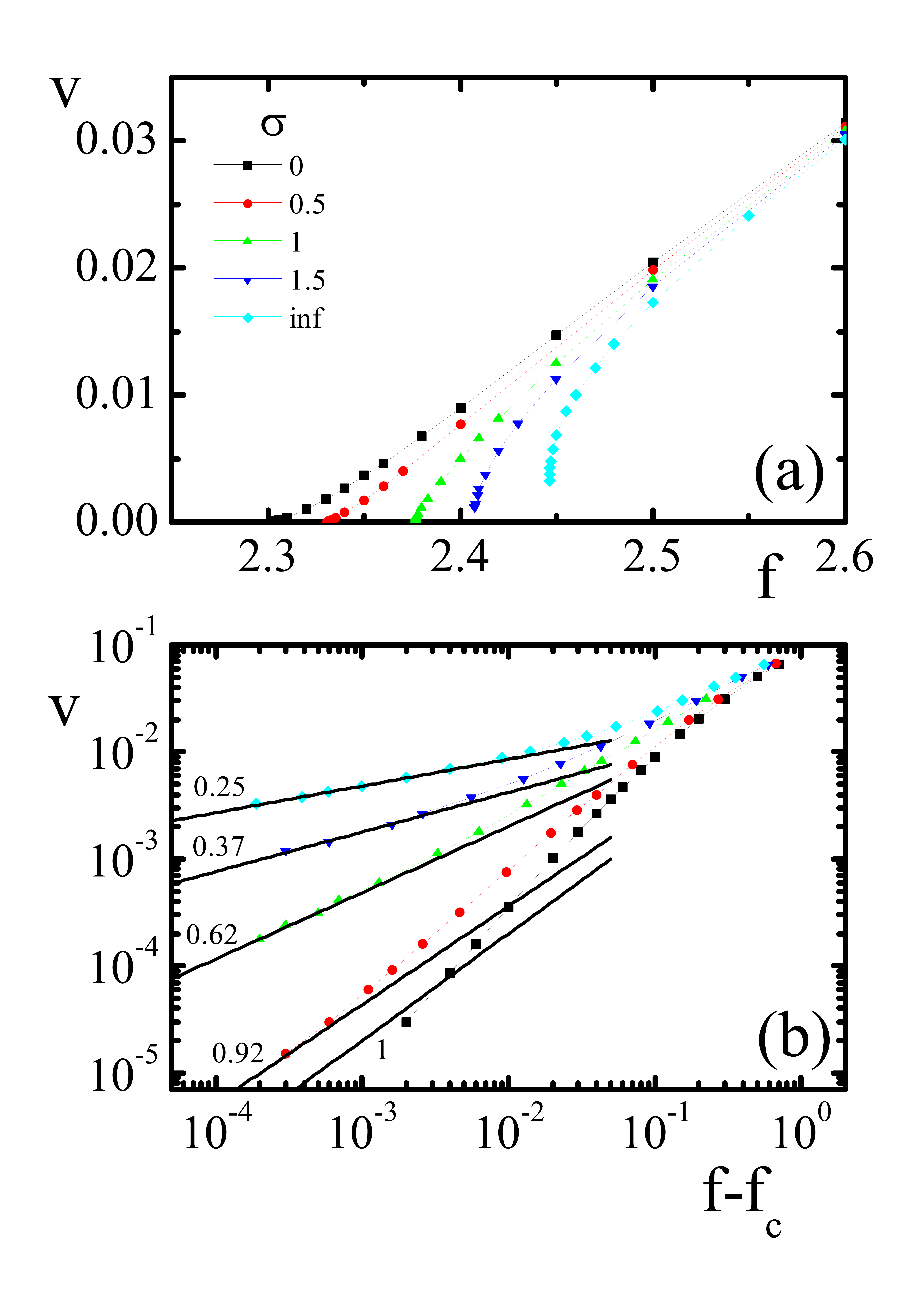}
\caption{(a) Flow curves of the discrete pinning potential model with rate transition $\lambda$ that depend on stress as $\lambda\sim (f-f^{\tt th})^{1/2}$, for different values of the long range interaction exponent $\sigma$. 
(b) Same data in logarithmic scale, using the value of $f_c$ previously used in Fig. \ref{lambda_cte}(b). The continuous straight lines have the same slope than in the previous Figure. 
\label{lambda_sqr}
}
\end{figure}

\begin{figure}
\includegraphics[width=7cm,clip=true]{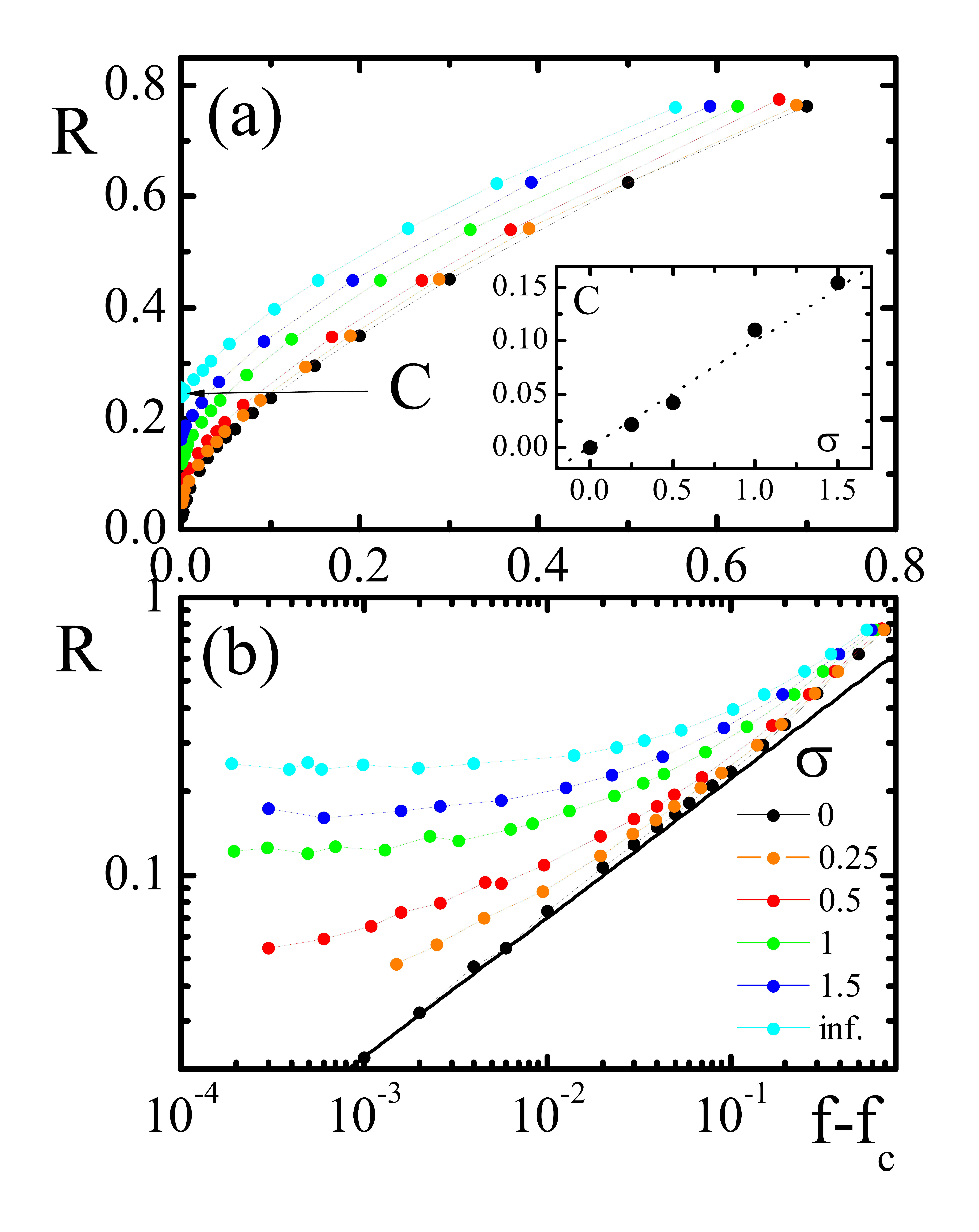}
\caption{Ratio $R$ between the velocities for variable and constant rate, as a function of the separation from the critical force. Linear (a) and logarithmic (b) scales. The limiting value $C$ for $f-f_c \to 0$ is plotted in the inset.
\label{cocientes}
}
\end{figure}

Fig. \ref{lambda_sqr}(a) displays the flow curves in the variable rate case, namely $\lambda\sim (f-f^{\tt th})^{1/2}$, mimicking the smooth continuous potential case. 
Comparing Figs. \ref{lambda_sqr}(a) and \ref{lambda_cte}(a) we observe again (as between Figs. \ref{f1}(a) and \ref{f1}(b)) a clear difference in the overall form of the curves with values of $\beta$ that look larger in Fig. \ref{lambda_sqr}(a) than in Fig. \ref{lambda_cte}(a). 
In order to quantify this difference in more detail it is necessary to look close to the critical force.
It was already mentioned that the great advantage of the discrete pinning model is that the values of critical force for the curves in Fig. \ref{lambda_sqr}(a)  are the same values as for the curves in Fig. \ref{lambda_cte}(a), which were already fitted to construct Fig. \ref{lambda_cte}(b). Using those values we construct the plot in Fig. \ref{lambda_sqr}(b).
We note that as far as $\sigma > 0$ the curves eventually reach the same exponent than in Fig. \ref{lambda_cte} when approaching the critical force. However, the force range in which this limiting behavior is obtained shrinks as $\sigma$ is reduced, and we are actually unable to observe it clearly once $\sigma\lesssim 0.5$. Our conclusion is that the critical region with the same $\beta$ as in the constant rate case remains finite for all $\sigma > 0$, but shrinks as $\sigma$ is reduced, vanishing for $\sigma\le 0$, where the mean field value $\beta=3/2$ is re-obtained.

An alternative way to look at the effect just described is to plot the ratio $R$ between the velocities for variable and constant rates (note that this has sense only in the present case in which the critical force is the same for the two different rates). The results contained in Fig. \ref{cocientes} show a remarkable systematic trend. As a function of $\Delta f\equiv f-f_c$, the velocity ratio
$R$ behaves as $R\sim C+A\Delta f ^{1/2}$. The value of $A$ is almost independent of the long range interaction exponent $\sigma$, but 
$C$ has a systematic dependence, reducing as $\sigma$ decreases, and vanishing at $\sigma=0$. For any $\sigma>0$ the finite value of $C$ implies the coincidence of the $\beta$ values for constant and variable rate. However, as $C$ decreases, the range to observe this coincidence decreases also, and for $\sigma=0$ (i.e., $C=0$) the value $\beta=3/2$ is obtained for variable rate, instead of the $\beta=1$ that is obtained for constant rate.

The estimation we have for $R$ allows to quantify the extent of the critical region. We can say that the critical region extends roughly up to the point where $C(\sigma) \sim A\Delta f^{1/2}$. This provides  $\Delta f^{\tt crit}\sim C(\sigma)^2$. As $C$ is observed to be roughly linear with $\sigma$, we finally obtain $\Delta f^{\tt crit}\sim \sigma^2$.

\section{Implications on the avalanche dynamics}

The discrete pinning potential model is also useful to make an accurate determination of avalanche statistics. 
In order to analyze it, we did simulations using a quasistatic algorithm, consisting in
progressively reducing the value of the external force $f$ in Eq.(\ref{fapp}) as dynamics proceeds. Namely, each time a site 
jump from $u_i$ to $u_i+\delta$, $f$ is reduced to $f-\delta/N$. In this way, $f$ eventually becomes 
lower than $f_c$ and the dynamic stops. At this point $f$ is increased up to the point in which one
site becomes unstable. This allows to analyze individual avalanches in the system at $f_c$.
In this way we generated large sequences of avalanches in the system, to which we can calculate duration $T$, size $S$ (calculated as the difference between $\sum_i u_i$ after and before the avalanche), and spatial extent $L$ (which is the number of sites that jumped during the avalanche). The statistical behavior of these three quantities allows to determine some important critical exponents of the transition. 
For instance, the (average) relation  between $S$ and $L$ allows to determine the roughness exponent $\zeta$ through $S\sim L^{d+\zeta}$. In addition the relation between $T$ and $L$ 
determines the dynamical exponent $z$: $T\sim L^z$. 

\begin{figure}
\includegraphics[width=7cm,clip=true]{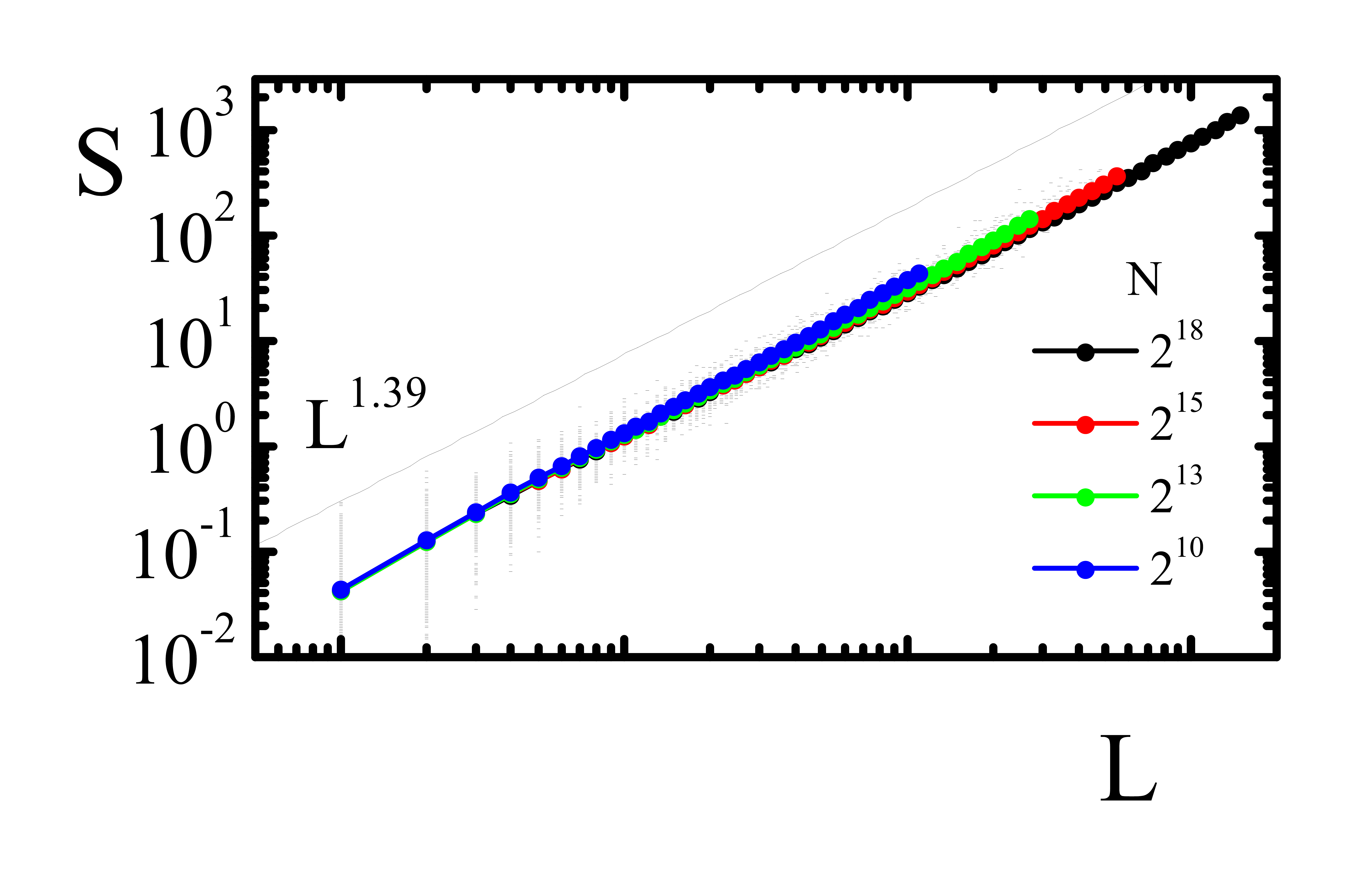}
\caption{Size $S$ vs spatial extent $L$ 
of individual avalanches (small dots) obtained in quasistatic simulations of the discrete pinning potential model, at $\sigma=1$. 
Averaging over small $L$ intervals the continuous lines are obtained for different system sizes, which allows to determine the value of the critical exponents $\zeta$. The value obtained $\zeta\simeq 0.39$ coincides with the one reported in the literature.
\label{SLT}
}
\end{figure}

In Fig. \ref{SLT}, $S$ vs $L$ is plotted for the ``standard" long range case $\sigma=1$, 
which is particularly relevant for propagating fractures, contact lines of liquids or ``magnetically charged'' 
domain walls~\cite{Zapperi1998}. 
By construction of the model, this plot is valid both for constant, and also for variable rates.
Although the individual data points are quite scattered, averaging over small intervals of $L$ allows to obtain a good estimation of $\zeta$. 
The value obtained ($\zeta \simeq 0.39$) perfectly coincides within the error bar with the value reported in the literature (see e.g. Ref. [~\onlinecite{Rosso2002}]). 

\begin{figure}
\includegraphics[width=7cm,clip=true]{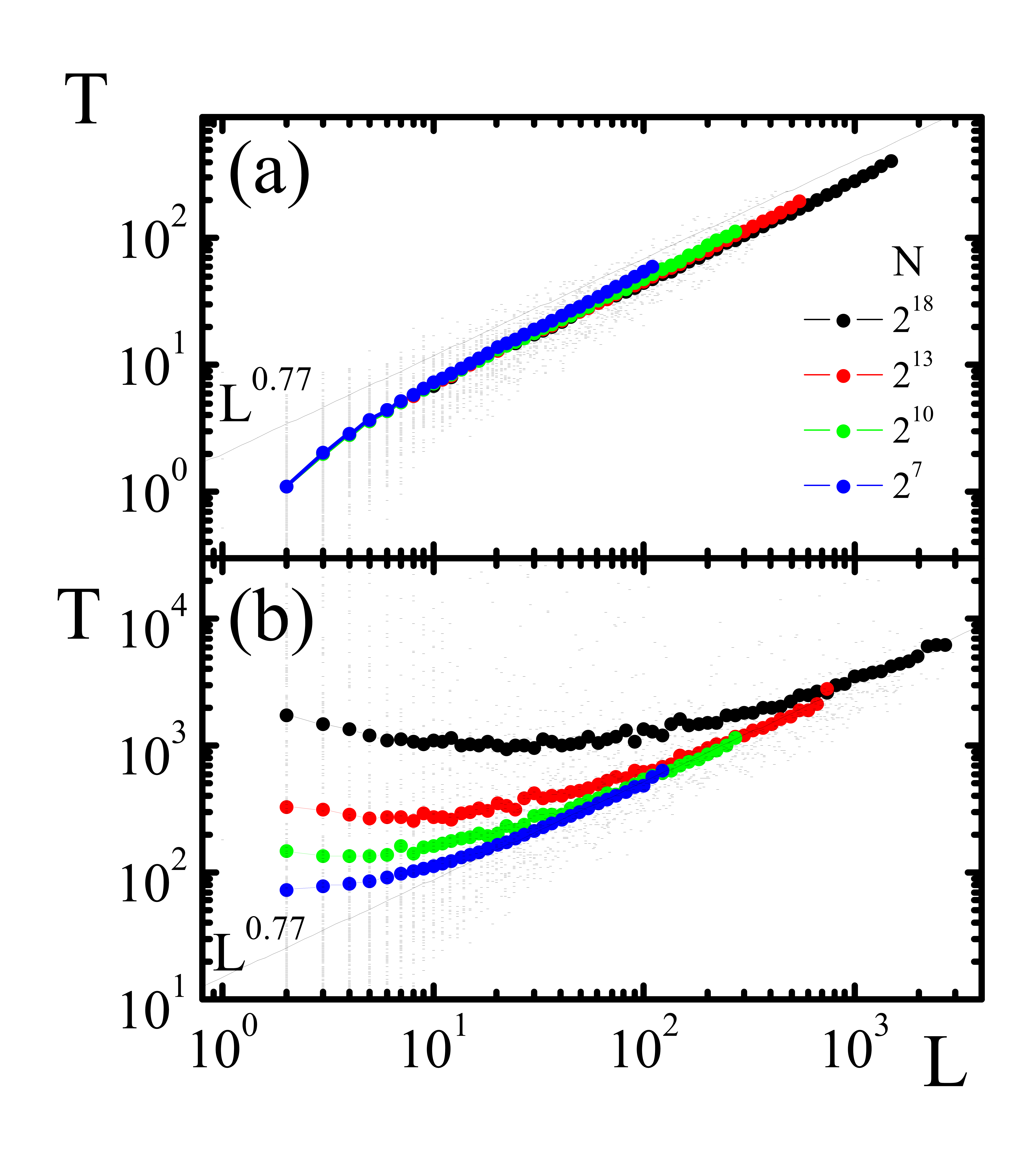}
\caption{Duration vs spatial extent 
of avalanches obtained in quasistatic simulations of the discrete pinning potential model with constant (a) and variable (b) transition rate, at $\sigma=1$.
\label{LT}
}
\end{figure}

We now focus on the duration $T$ vs spatial extent $L$ relation, determining the dynamical exponent $z$ as $T\sim L^z$. 
This result depends on the form of the rates.
For constant rate the result is shown in Fig. \ref{LT}(a), and is consistent with the expected value of $z$, namely $z\simeq 0.77$~\cite{Duemmer2007,Ramanathan1998}. The results in Figs. \ref{SLT} and \ref{LT}(a) further support the claim that the constant rate discrete pinning potential is a realization of the cuspy continuous potential case.

The results for $T$ vs $L$ for the variable rate case are presented in Fig.  \ref{LT}(b). 
The first thing that is observed is that data for individual avalanches (small dots) are much more scattered compared to Fig. \ref{LT}(a). This behavior is clearly related to the variable rate: whereas for constant rate the avalanche duration is at most of the order of avalanche size, for variable rate even small avalanches can last for quite long, as a single site may take a very long time to be activated if it is only slightly above the local critical force.
The next observation in Fig. \ref{LT}(b) is that a relation $T \sim L^z$ with $z$ being the same exponent as in Fig. \ref{LT}(a) is obtained, but only for sufficiently large avalanches. This is in fact consistent with our view that the critical region for the 
variable rate case is much smaller than for the constant rate case, and only large avalanches display the correct critical $T$ vs $L$
dependence. For small avalanches this dependence deviates towards larger values of $T$. 
Interestingly, cracks experiments ~\cite{Bonamy2008} also show an``excess duration'' for small avalanches, suggesting the non-universal effect of smooth microscopic disorder or ``variable rate'' local instabilities.  
Note also that for these small avalanches the typical duration depends also on the system size $N$, an effect that was already discussed in Ref.\onlinecite{Aguirre2018} in the context of the yielding transition.

\begin{figure}
\includegraphics[width=7cm,clip=true]{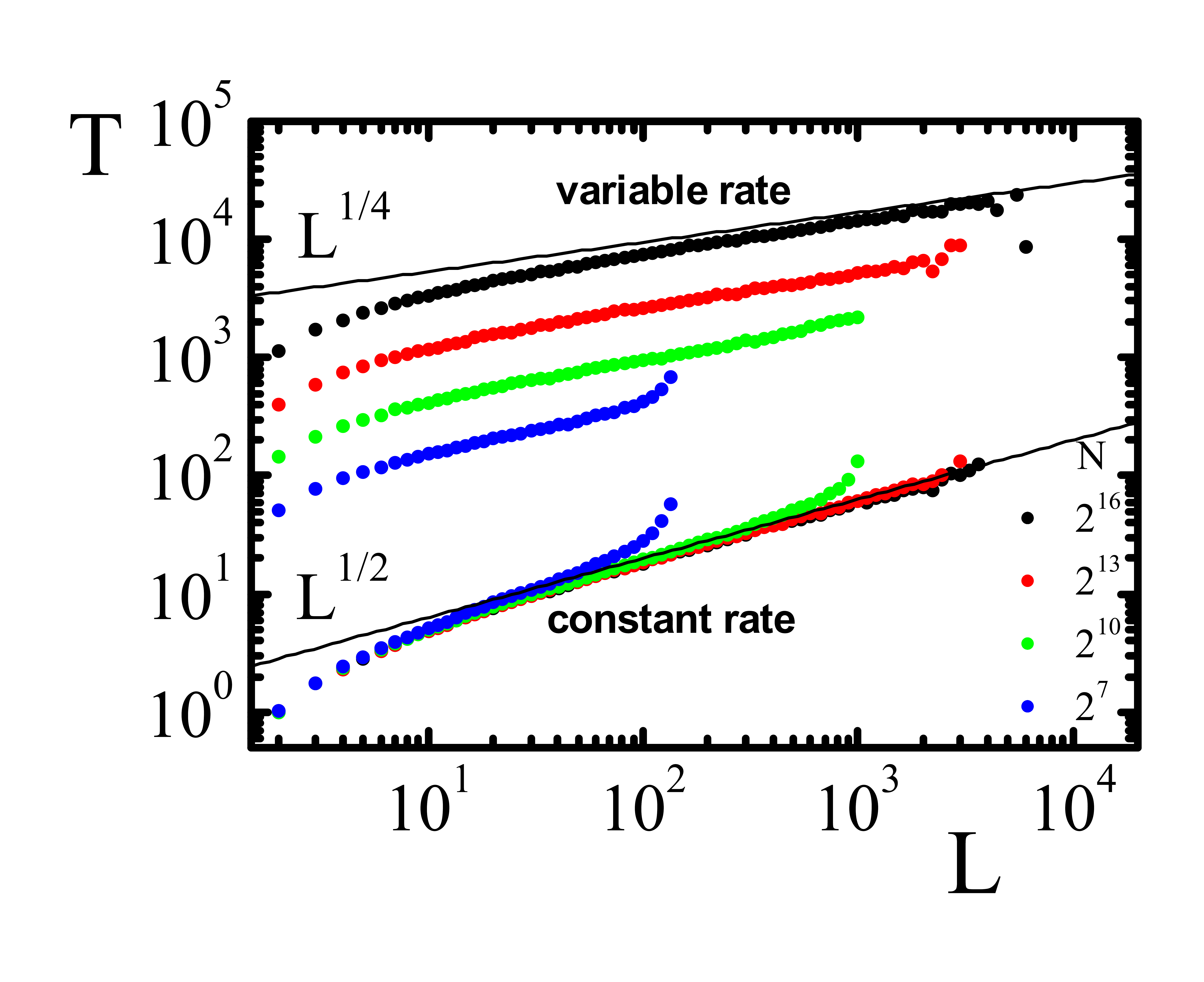}
\caption{Duration vs spatial extent 
of avalanches obtained in quasistatic simulations of the discrete pinning potential model, in a mean field situation ($\sigma=0$).
For constant transition rate the expected result $z=1/2$ is obtained, and there are no dependences on system size besides the appearance of progressively larger avalanches as system size increases. For variable transition rate the value $z=1/4$ is obtained and in addition, a global dependence on system size of the form $N^{1/2}$ is observed.
\label{mf}
}
\end{figure}

Since the results for the flow curve support the idea that the size of the critical region shrinks to zero as $\sigma\to 0$, the 
question naturally arises of what is the nature of the $T$ vs $L$ dependence and the value of $z$ in this limit, for constant and variable rates.
To address this point we generated avalanches in a mean field case ($\sigma=0$), and plot the results in Fig. \ref{mf}, both for constant and variable rate, for system of different sizes.
The results for constant rate are consistent with the standard value of $z$ in mean field, namely $z=1/2$. However, results for variable rate sharply deviate of this behavior. We obtain a value of $z=1/4$ instead, and also an overall dependence on the system size, in such a way that we can write $T\sim L^{1/4}N^{1/2}$. This last relation can in fact be obtained analytically~\footnote{I. Fern\'andez Aguirre, A. Rosso, and E. A. Jagla, unpublished.}. 

\section{Conclusions}
\label{sec:conclusions}
\begin{figure}
\includegraphics[width=\columnwidth,clip=true]{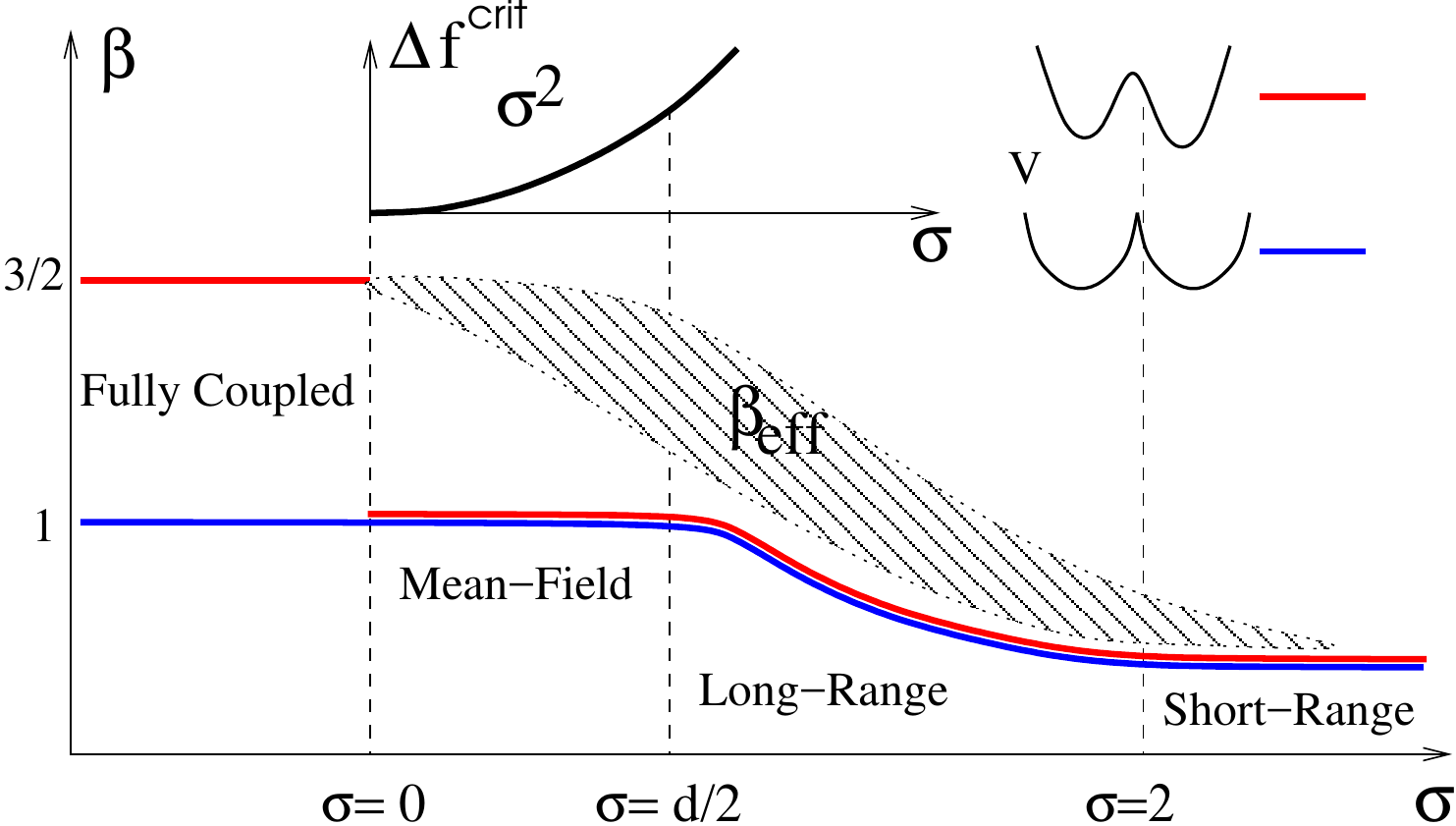}
\caption{Schematic illustration of the answer to the puzzle posed in Fig. \ref{fig:question}.
The extent of the critical region for smooth potentials (measured by $\Delta f^{\tt crit}$) reduces continuously to zero as $\sigma=0$ is approached. In the region $\sigma\le 0$ the non-universal critical exponent $\beta=3/2$ is observed. This raises the possibility (in experiments or numerical simulations) to observe strong corrections to scaling in the smooth potential case, specially at low values of $\sigma$, that may induce to adjust  effective values of the $\beta$ exponent (roughly indicated by the hatched region), 
that are expected to be larger than the true values.
\label{fig:answer}
}
\end{figure}

In Fig.\ref{fig:answer} we summarize the picture 
that emerges from our results. The 
peculiar breakdown of universality in the 
$\sigma \to 0$ limit is explained in terms of a vanishing critical 
region for smooth potentials (whenever disorder is strong enough 
to have a finite $f_c$ in such limit).
\begin{figure}
\includegraphics[width=\columnwidth,clip=true]{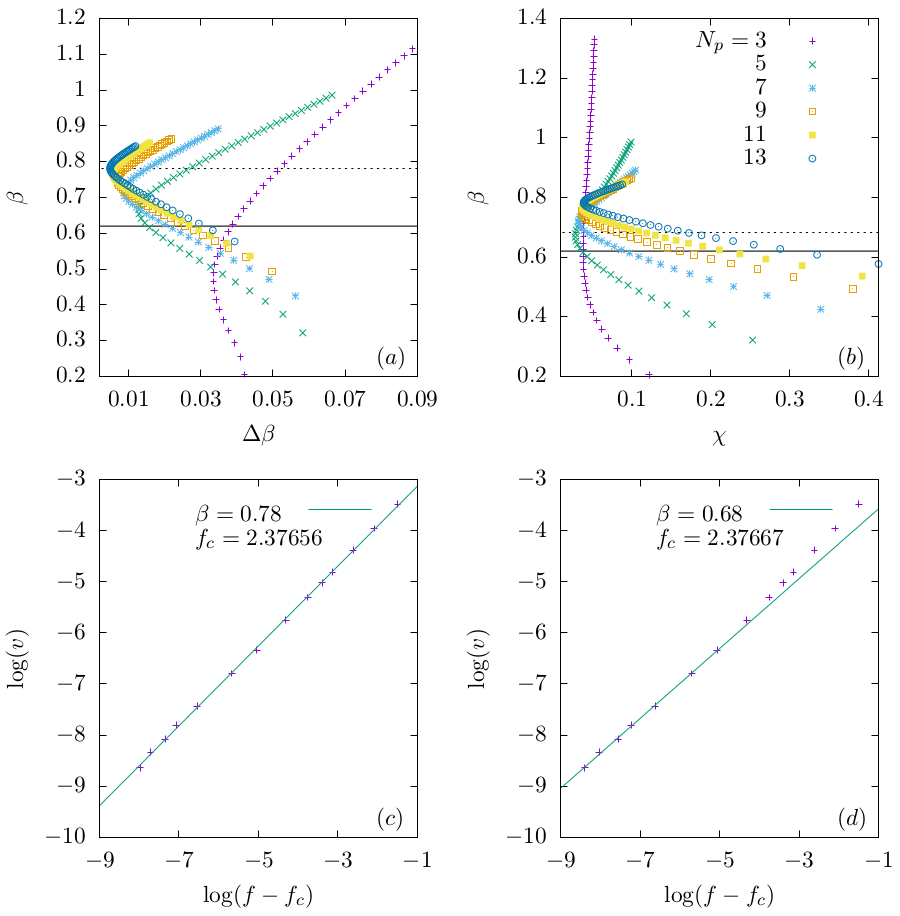}
\caption{Estimations of $\beta$ and $f_c$ from the raw data points of 
Fig.\ref{lambda_sqr} for $\sigma=1$ corresponding 
to the smooth microscopic pinning or variable jump rate 
case. The best pairs 
$f_c$ and $\beta$ are obtained from least squares fits to 
$\log v = \beta \log(f-f_c)+\text{cte}$ varying the number 
$N_p$ of points considered from the lowest $f$. From (a) we 
choose the values that minimize the fit parameter error 
$\Delta \beta$, and obtain the fit shown in (c) 
(the effective $\beta$ is shown by a dashed line in (a))
From (b) we 
choose the values that minimize the standard deviation $\chi$  
of the fit and obtain the fit shown in (d) (the effective 
$\beta$ is shown by a dashed line in (b)). 
The obtained values must be compared with the asymptotic ones 
$\beta=0.62$ (shown with a solid line in (a) and (b)), 
and $f_c=2.37670$. 
\label{figexp}
}
\end{figure}
We argue that the vanishing of the critical region has 
practical implications for the analysis of long-range 
depinning. Since the 
fully-coupled model has a $\beta=3/2$ exponent for 
smooth potentials, larger than the universal 
$\beta=1$ for cuspy potentials, effective 
values $3/2>\beta_{\tt eff}>1$ are plausible to 
be observed for $\sigma \gtrsim 0$, 
but also for even larger $\sigma$ we expect 
an excess, i.e. 
$\beta_{\tt eff}(\sigma) > \beta(\sigma)$.
The difference between the effective and the right exponents 
are found to be more important for the dynamical exponents, $\beta$ 
and $z$, than for the geometric exponents, $\zeta$ and $\nu$.
From a simple microscopic model we have shown that this is  
related to the competition between the characteristic time 
$\tau_1$ associated to single particle instabilities,
with the time associated to collective instabilities, 
$\tau \sim (f-f_c)^{-z\nu}$, which is roughly controlled by the 
number of active particles involved in the spreading of 
correlations at lengths $l\sim (f-f_c)^{-\nu}$.
This competition is made more clear when we analyze 
avalanche dynamics in the quasistatic limit.

To illustrate the kind of effects we can expect from 
the non-universal corrections to scaling arising from 
smooth microscopic pinning potentials, 
in Fig.\ref{figexp} we show estimations of $\beta$ and $f_c$ 
from the raw data points of Fig.\ref{lambda_sqr} for $\sigma=1$ 
corresponding to the variable jump rate 
case discussed in Section \ref{sec:discrete}.
We use two fitting methods which are often used in the 
literature. The best pairs 
$f_c$ and $\beta$ are obtained from least squares fits to 
$\log v = \beta \log(f-f_c)+\text{cte}$ varying the number 
$N_p$ of points considered, starting from the three lowest values of 
$f$. For each case we slowly decrease $f_c$ from the lowest value of 
$f$. In Fig.\ref{figexp}(a) we show that for each $N_p$ (labels for 
each $N_p$ are shared with Fig.\ref{figexp}(b))  
the fit parameter error $\Delta \beta$ displays a minimum for a given value of $f_c$. 
If we choose the values corresponding to this minimum we obtain 
the fairly good fit of the data shown in Fig.\ref{figexp}(c). 
If instead we choose the values corresponding to the minimum of
the standard deviation $\chi$ of the fit we obtain 
the fit shown in Fig.\ref{figexp}(c) (note that in this case, the optimum $N_p$ and $f_c$ 
are different than with the previous criteria). 
The obtained values must be compared with those obtained from the more robust 
constant rate simulations (Fig. \ref{lambda_cte})
$\beta=0.62$, and $f_c=2.37670$. It is worth noting that 
in either case the effective $\beta$ is larger than its true asymptotic value, 
which is accurately obtained by using the constant rate discrete model.

Our results motivate a reexamination of the empirical experimental and numerical 
(smooth potential) long range depinning data analysis. 
In Ref.\cite{Ponson2009} the depinning exponent $\beta\approx 0.8$ 
was directly measured for cracks propagating in an elastic inhomogeneous material.
A less direct estimate, also for propagating cracks, 
can be obtained from the experimental results for the avalanche 
duration exponent $\gamma=1.67$ reported in Ref.~\cite{Laurson2013}.
Using that $\gamma \approx \beta + \zeta/(1+\zeta)$ and 
assuming $\zeta=0.39$ ~\cite{Rosso2002,Duemmer2007} we get
$\beta \approx 0.72$.
Both values appear to be larger than the ones predicted for the universality class of 
one dimensional elastic interfaces with $\sigma=1$ long-range elastic couplings 
and uncorrelated isotropic disorder, where
$\beta \approx 0.63$ ~\cite{Duemmer2007,Laurson2013} and $\beta \approx 0.68$ 
~\cite{Ramanathan1998} were found numerically using 
``cuspy'' or cellular automata lattice models. 
One can thus argue that the excess in the effective  
value of $\beta$ may arise, in part, from the 
strong corrections to scaling we expect for 
long-range depinning with smooth microscopic pinning potentials, 
due to the vanishing of the critical region approaching 
the fully-coupled limit.

\begin{acknowledgments}
We thank A. Rosso for enlightening discussions.
A.B.K. acknowledges hospitality at LPTMS-Universit\'e Paris-Sud.
This work was partly supported by grants PICT2016-0069/FONCyT 
and UNCuyo C017, from Argentina.
\end{acknowledgments}

\appendix

\section{Results in higher dimensions}

In addition to considering a long range interaction in a one dimensional system, there is a second standard way to move towards the mean field limit. This is to consider the short range depinning problem in progressively larger number of dimensions.
For short range interactions, the critical dimension of the depinning problem is $d_c=4$, i.e, for $d>d_c$ we expect mean field critical exponents, in particular $\beta=1$. The short range case in $d_c=4$ corresponds to the $\sigma=1/2$ case of a 1D system.
It is then natural to ask if increasing the dimension $d$ we observe the same trends we observe by decreasing the exponent $\sigma$, particularly in the flow curves.

\begin{figure}
\includegraphics[width=7cm,clip=true]{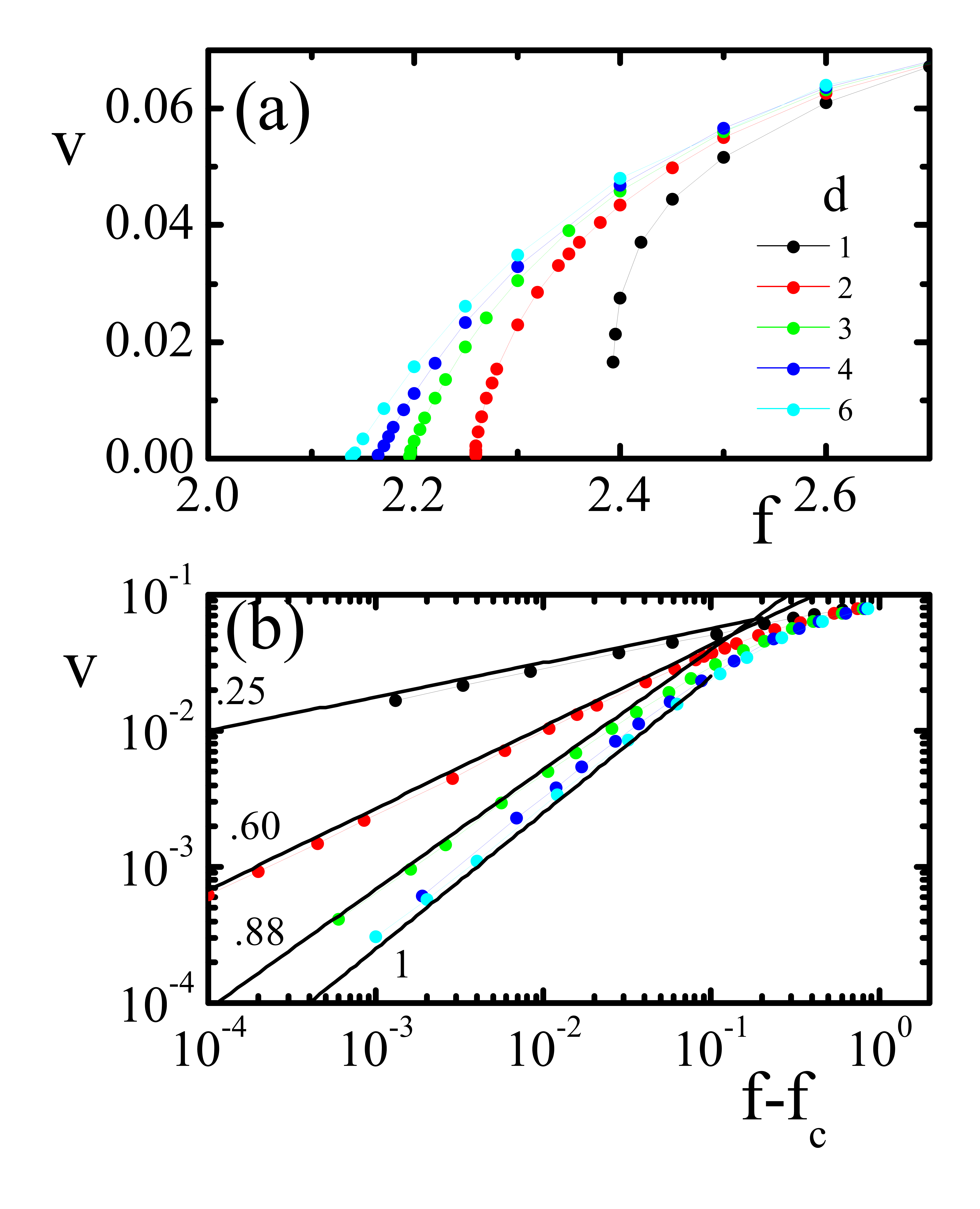}
\caption{Velocity as a function of force for interfaces with short-range elasticity and 
constant transition rates potentials for interfaces of different dimensionality $d$ 
in linear (a) and logarithmic (b) scale. The fitted values of $\beta$ are indicated.
\label{234da}
}
\end{figure}

\begin{figure}
\includegraphics[width=7cm,clip=true]{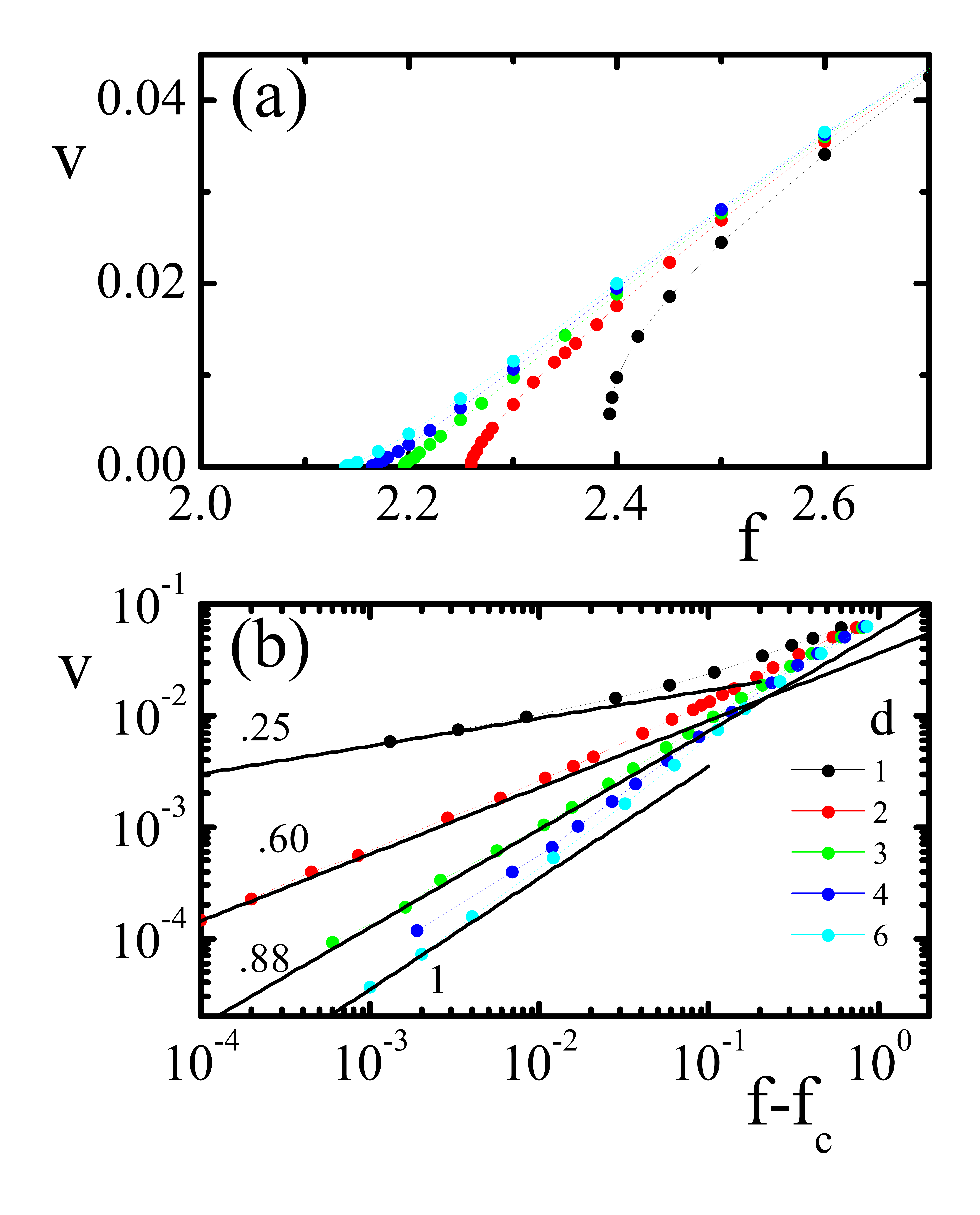}
\caption{Same as previous figure for variable transition rates.
\label{234db}
}
\end{figure}

\begin{figure}
\includegraphics[width=7cm,clip=true]{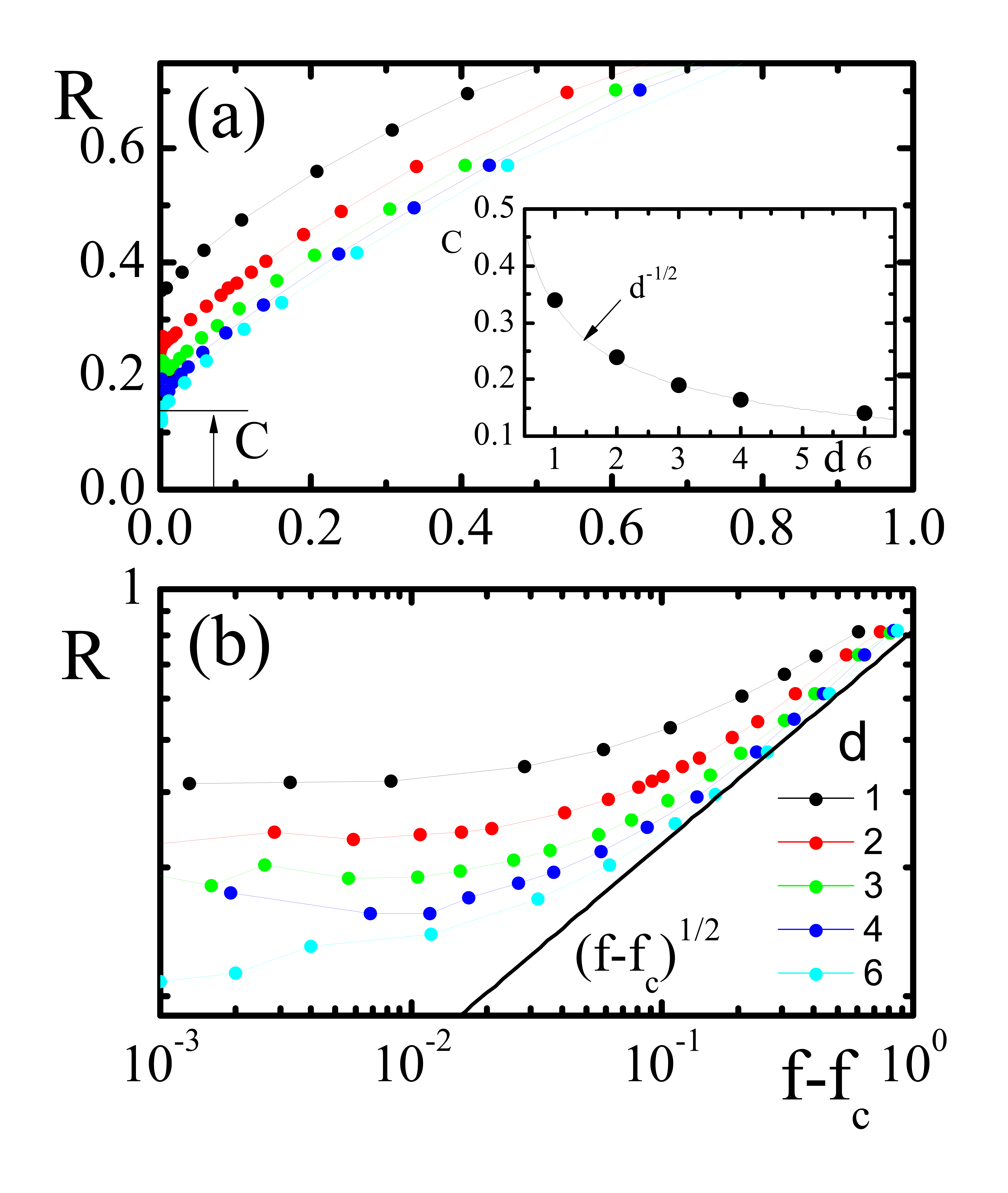}
\caption{Ratio between the velocity for variable and constant transition rates. We observe the same trend as it was obtained in 1D long range interacting systems as a function of $\sigma$ (Fig. \ref{cocientes}). The value of $C$ decreases with $d$ as $d^{-1/2}$
\label{234d}
}
\end{figure}

We have found that the answer to this question is affirmative. The relevant results are contained in Figs. \ref{234da}, \ref{234db}, and \ref{234d} (which should be directly compared to Figs. \ref{lambda_cte}, \ref{lambda_sqr}, and \ref{cocientes}). There we present simulations of the discrete pinning potential model,  in different number of spatial dimensions, namely $d=1$, 2, 3, 4 and 6, with interactions only among nearest neighbor sites (corresponding to $\sigma\to\infty$). In order to compare different dimensions more easily, and to have a well defined limit as $d\to \infty$, here the elastic interaction $G_{ij}$ (in Eq. \ref{eqmotion}) is normalized differently, namely we take the value of  $G_{ij}$ for neighbor sites as $2/d$.
The trend we observe as dimension is increased is equivalent to what we have obtained as $\sigma$ is reduced, in one dimension: For constant rate a robust critical region is obtained and the value of $\beta$ increases with the number of dimensions, reaching the mean field value $\beta=1$ for $d=4$. For variable rates the extent of the critical region is smaller, and is reduced as $d$ increases. Note however that the critical region does not vanish at the upper critical dimension $d=4$: our data are consistent with the critical region vanishing only as $d\to\infty$ where the mean field value ($\beta=3/2$) is fully observed. As in the case of varying $\sigma$, here we can estimate the extent of the critical region as a function of $d$ (see Fig. \ref{234d}) and the result is $\Delta f^{\tt crit}\sim d^{-1}$.

\bibliography{biblio}

\begin{thebibliography}{74}%
\makeatletter
\providecommand \@ifxundefined [1]{%
 \@ifx{#1\undefined}
}%
\providecommand \@ifnum [1]{%
 \ifnum #1\expandafter \@firstoftwo
 \else \expandafter \@secondoftwo
 \fi
}%
\providecommand \@ifx [1]{%
 \ifx #1\expandafter \@firstoftwo
 \else \expandafter \@secondoftwo
 \fi
}%
\providecommand \natexlab [1]{#1}%
\providecommand \enquote  [1]{``#1''}%
\providecommand \bibnamefont  [1]{#1}%
\providecommand \bibfnamefont [1]{#1}%
\providecommand \citenamefont [1]{#1}%
\providecommand \href@noop [0]{\@secondoftwo}%
\providecommand \href [0]{\begingroup \@sanitize@url \@href}%
\providecommand \@href[1]{\@@startlink{#1}\@@href}%
\providecommand \@@href[1]{\endgroup#1\@@endlink}%
\providecommand \@sanitize@url [0]{\catcode `\\12\catcode `\$12\catcode
  `\&12\catcode `\#12\catcode `\^12\catcode `\_12\catcode `\%12\relax}%
\providecommand \@@startlink[1]{}%
\providecommand \@@endlink[0]{}%
\providecommand \url  [0]{\begingroup\@sanitize@url \@url }%
\providecommand \@url [1]{\endgroup\@href {#1}{\urlprefix }}%
\providecommand \urlprefix  [0]{URL }%
\providecommand \Eprint [0]{\href }%
\providecommand \doibase [0]{http://dx.doi.org/}%
\providecommand \selectlanguage [0]{\@gobble}%
\providecommand \bibinfo  [0]{\@secondoftwo}%
\providecommand \bibfield  [0]{\@secondoftwo}%
\providecommand \translation [1]{[#1]}%
\providecommand \BibitemOpen [0]{}%
\providecommand \bibitemStop [0]{}%
\providecommand \bibitemNoStop [0]{.\EOS\space}%
\providecommand \EOS [0]{\spacefactor3000\relax}%
\providecommand \BibitemShut  [1]{\csname bibitem#1\endcsname}%
\let\auto@bib@innerbib\@empty
\bibitem [{\citenamefont {Durin}\ and\ \citenamefont
  {Zapperi}(2006)}]{Durin2006}%
  \BibitemOpen
  \bibfield  {author} {\bibinfo {author} {\bibfnamefont {G.}~\bibnamefont
  {Durin}}\ and\ \bibinfo {author} {\bibfnamefont {S.}~\bibnamefont
  {Zapperi}},\ }in\ \href@noop {} {\emph {\bibinfo {booktitle} {The Science of
  Hysteresis: Physical Modeling, Micromagnetics and Magnetization Dynamics,
  vol. II}}},\ \bibinfo {editor} {edited by\ \bibinfo {editor} {\bibfnamefont
  {I.~D.}\ \bibnamefont {Mayergoyz}}}\ (\bibinfo  {publisher} {Cambridge
  University Press},\ \bibinfo {year} {2006})\ Chap.~\bibinfo {chapter}
  {10}\BibitemShut {NoStop}%
\bibitem [{\citenamefont {Ferré}\ \emph {et~al.}(2013)\citenamefont {Ferré},
  \citenamefont {Metaxas}, \citenamefont {Mougin}, \citenamefont {Jamet},
  \citenamefont {Gorchon},\ and\ \citenamefont {Jeudy}}]{Ferre2013}%
  \BibitemOpen
  \bibfield  {author} {\bibinfo {author} {\bibfnamefont {J.}~\bibnamefont
  {Ferré}}, \bibinfo {author} {\bibfnamefont {P.~J.}\ \bibnamefont {Metaxas}},
  \bibinfo {author} {\bibfnamefont {A.}~\bibnamefont {Mougin}}, \bibinfo
  {author} {\bibfnamefont {J.-P.}\ \bibnamefont {Jamet}}, \bibinfo {author}
  {\bibfnamefont {J.}~\bibnamefont {Gorchon}}, \ and\ \bibinfo {author}
  {\bibfnamefont {V.}~\bibnamefont {Jeudy}},\ }\href {\doibase
  https://doi.org/10.1016/j.crhy.2013.08.001} {\bibfield  {journal} {\bibinfo
  {journal} {Comptes Rendus Physique}\ }\textbf {\bibinfo {volume} {14}},\
  \bibinfo {pages} {651 } (\bibinfo {year} {2013})},\ \bibinfo {note}
  {disordered systems / Systèmes désordonnés}\BibitemShut {NoStop}%
\bibitem [{\citenamefont {Durin}\ \emph {et~al.}(2016)\citenamefont {Durin},
  \citenamefont {Bohn}, \citenamefont {Corr\^ea}, \citenamefont {Sommer},
  \citenamefont {Le~Doussal},\ and\ \citenamefont {Wiese}}]{Durin2016}%
  \BibitemOpen
  \bibfield  {author} {\bibinfo {author} {\bibfnamefont {G.}~\bibnamefont
  {Durin}}, \bibinfo {author} {\bibfnamefont {F.}~\bibnamefont {Bohn}},
  \bibinfo {author} {\bibfnamefont {M.~A.}\ \bibnamefont {Corr\^ea}}, \bibinfo
  {author} {\bibfnamefont {R.~L.}\ \bibnamefont {Sommer}}, \bibinfo {author}
  {\bibfnamefont {P.}~\bibnamefont {Le~Doussal}}, \ and\ \bibinfo {author}
  {\bibfnamefont {K.~J.}\ \bibnamefont {Wiese}},\ }\href {\doibase
  10.1103/PhysRevLett.117.087201} {\bibfield  {journal} {\bibinfo  {journal}
  {Phys. Rev. Lett.}\ }\textbf {\bibinfo {volume} {117}},\ \bibinfo {pages}
  {087201} (\bibinfo {year} {2016})}\BibitemShut {NoStop}%
\bibitem [{\citenamefont {Kleemann}(2007)}]{Kleemann2007}%
  \BibitemOpen
  \bibfield  {author} {\bibinfo {author} {\bibfnamefont {W.}~\bibnamefont
  {Kleemann}},\ }\href {\doibase 10.1146/annurev.matsci.37.052506.084243}
  {\bibfield  {journal} {\bibinfo  {journal} {Annual Review of Materials
  Research}\ }\textbf {\bibinfo {volume} {37}},\ \bibinfo {pages} {415}
  (\bibinfo {year} {2007})},\ \Eprint
  {http://arxiv.org/abs/https://doi.org/10.1146/annurev.matsci.37.052506.084243}
  {https://doi.org/10.1146/annurev.matsci.37.052506.084243} \BibitemShut
  {NoStop}%
\bibitem [{\citenamefont {Paruch}\ and\ \citenamefont
  {Guyonnet}(2013)}]{Paruch2013}%
  \BibitemOpen
  \bibfield  {author} {\bibinfo {author} {\bibfnamefont {P.}~\bibnamefont
  {Paruch}}\ and\ \bibinfo {author} {\bibfnamefont {J.}~\bibnamefont
  {Guyonnet}},\ }\href {\doibase https://doi.org/10.1016/j.crhy.2013.08.004}
  {\bibfield  {journal} {\bibinfo  {journal} {Comptes Rendus Physique}\
  }\textbf {\bibinfo {volume} {14}},\ \bibinfo {pages} {667 } (\bibinfo {year}
  {2013})},\ \bibinfo {note} {disordered systems / Systèmes
  désordonnés}\BibitemShut {NoStop}%
\bibitem [{\citenamefont {Bonamy}\ \emph {et~al.}(2008)\citenamefont {Bonamy},
  \citenamefont {Santucci},\ and\ \citenamefont {Ponson}}]{Bonamy2008}%
  \BibitemOpen
  \bibfield  {author} {\bibinfo {author} {\bibfnamefont {D.}~\bibnamefont
  {Bonamy}}, \bibinfo {author} {\bibfnamefont {S.}~\bibnamefont {Santucci}}, \
  and\ \bibinfo {author} {\bibfnamefont {L.}~\bibnamefont {Ponson}},\ }\href
  {\doibase 10.1103/PhysRevLett.101.045501} {\bibfield  {journal} {\bibinfo
  {journal} {Phys. Rev. Lett.}\ }\textbf {\bibinfo {volume} {101}},\ \bibinfo
  {pages} {045501} (\bibinfo {year} {2008})}\BibitemShut {NoStop}%
\bibitem [{\citenamefont {Ponson}(2009)}]{Ponson2009}%
  \BibitemOpen
  \bibfield  {author} {\bibinfo {author} {\bibfnamefont {L.}~\bibnamefont
  {Ponson}},\ }\href {\doibase 10.1103/PhysRevLett.103.055501} {\bibfield
  {journal} {\bibinfo  {journal} {Phys. Rev. Lett.}\ }\textbf {\bibinfo
  {volume} {103}},\ \bibinfo {pages} {055501} (\bibinfo {year}
  {2009})}\BibitemShut {NoStop}%
\bibitem [{\citenamefont {Moulinet}\ \emph {et~al.}(2004)\citenamefont
  {Moulinet}, \citenamefont {Rosso}, \citenamefont {Krauth},\ and\
  \citenamefont {Rolley}}]{moulinet2004}%
  \BibitemOpen
  \bibfield  {author} {\bibinfo {author} {\bibfnamefont {S.}~\bibnamefont
  {Moulinet}}, \bibinfo {author} {\bibfnamefont {A.}~\bibnamefont {Rosso}},
  \bibinfo {author} {\bibfnamefont {W.}~\bibnamefont {Krauth}}, \ and\ \bibinfo
  {author} {\bibfnamefont {E.}~\bibnamefont {Rolley}},\ }\href {\doibase
  10.1103/PhysRevE.69.035103} {\bibfield  {journal} {\bibinfo  {journal} {Phys.
  Rev. E}\ }\textbf {\bibinfo {volume} {69}},\ \bibinfo {pages} {035103}
  (\bibinfo {year} {2004})}\BibitemShut {NoStop}%
\bibitem [{\citenamefont {Doussal}\ \emph {et~al.}(2009)\citenamefont
  {Doussal}, \citenamefont {Wiese}, \citenamefont {Moulinet},\ and\
  \citenamefont {Rolley}}]{ledoussal2009}%
  \BibitemOpen
  \bibfield  {author} {\bibinfo {author} {\bibfnamefont {P.~L.}\ \bibnamefont
  {Doussal}}, \bibinfo {author} {\bibfnamefont {K.~J.}\ \bibnamefont {Wiese}},
  \bibinfo {author} {\bibfnamefont {S.}~\bibnamefont {Moulinet}}, \ and\
  \bibinfo {author} {\bibfnamefont {E.}~\bibnamefont {Rolley}},\ }\href
  {http://stacks.iop.org/0295-5075/87/i=5/a=56001} {\bibfield  {journal}
  {\bibinfo  {journal} {EPL (Europhysics Letters)}\ }\textbf {\bibinfo {volume}
  {87}},\ \bibinfo {pages} {56001} (\bibinfo {year} {2009})}\BibitemShut
  {NoStop}%
\bibitem [{\citenamefont {Planet}\ \emph {et~al.}(2009)\citenamefont {Planet},
  \citenamefont {Santucci},\ and\ \citenamefont {Ort\'{\i}n}}]{Planet2009}%
  \BibitemOpen
  \bibfield  {author} {\bibinfo {author} {\bibfnamefont {R.}~\bibnamefont
  {Planet}}, \bibinfo {author} {\bibfnamefont {S.}~\bibnamefont {Santucci}}, \
  and\ \bibinfo {author} {\bibfnamefont {J.}~\bibnamefont {Ort\'{\i}n}},\
  }\href {\doibase 10.1103/PhysRevLett.102.094502} {\bibfield  {journal}
  {\bibinfo  {journal} {Phys. Rev. Lett.}\ }\textbf {\bibinfo {volume} {102}},\
  \bibinfo {pages} {094502} (\bibinfo {year} {2009})}\BibitemShut {NoStop}%
\bibitem [{\citenamefont {Atis}\ \emph {et~al.}(2015)\citenamefont {Atis},
  \citenamefont {Dubey}, \citenamefont {Salin}, \citenamefont {Talon},
  \citenamefont {Le~Doussal},\ and\ \citenamefont {Wiese}}]{Atis2015}%
  \BibitemOpen
  \bibfield  {author} {\bibinfo {author} {\bibfnamefont {S.}~\bibnamefont
  {Atis}}, \bibinfo {author} {\bibfnamefont {A.~K.}\ \bibnamefont {Dubey}},
  \bibinfo {author} {\bibfnamefont {D.}~\bibnamefont {Salin}}, \bibinfo
  {author} {\bibfnamefont {L.}~\bibnamefont {Talon}}, \bibinfo {author}
  {\bibfnamefont {P.}~\bibnamefont {Le~Doussal}}, \ and\ \bibinfo {author}
  {\bibfnamefont {K.~J.}\ \bibnamefont {Wiese}},\ }\href {\doibase
  10.1103/PhysRevLett.114.234502} {\bibfield  {journal} {\bibinfo  {journal}
  {Phys. Rev. Lett.}\ }\textbf {\bibinfo {volume} {114}},\ \bibinfo {pages}
  {234502} (\bibinfo {year} {2015})}\BibitemShut {NoStop}%
\bibitem [{\citenamefont {Bayart}\ \emph {et~al.}(2015)\citenamefont {Bayart},
  \citenamefont {Svetlizky},\ and\ \citenamefont {Fineberg}}]{Bayart2015}%
  \BibitemOpen
  \bibfield  {author} {\bibinfo {author} {\bibfnamefont {E.}~\bibnamefont
  {Bayart}}, \bibinfo {author} {\bibfnamefont {I.}~\bibnamefont {Svetlizky}}, \
  and\ \bibinfo {author} {\bibfnamefont {J.}~\bibnamefont {Fineberg}},\ }\href
  {http://dx.doi.org/10.1038/nphys3539} {\bibfield  {journal} {\bibinfo
  {journal} {Nature Physics}\ }\textbf {\bibinfo {volume} {12}},\ \bibinfo
  {pages} {166 EP } (\bibinfo {year} {2015})}\BibitemShut {NoStop}%
\bibitem [{\citenamefont {{Nicolas}}\ \emph {et~al.}(2017)\citenamefont
  {{Nicolas}}, \citenamefont {{Ferrero}}, \citenamefont {{Martens}},\ and\
  \citenamefont {{Barrat}}}]{Nicolas2017}%
  \BibitemOpen
  \bibfield  {author} {\bibinfo {author} {\bibfnamefont {A.}~\bibnamefont
  {{Nicolas}}}, \bibinfo {author} {\bibfnamefont {E.~E.}\ \bibnamefont
  {{Ferrero}}}, \bibinfo {author} {\bibfnamefont {K.}~\bibnamefont
  {{Martens}}}, \ and\ \bibinfo {author} {\bibfnamefont {J.-L.}\ \bibnamefont
  {{Barrat}}},\ }\href@noop {} {\bibfield  {journal} {\bibinfo  {journal}
  {ArXiv e-prints}\ } (\bibinfo {year} {2017})},\ \Eprint
  {http://arxiv.org/abs/1708.09194} {arXiv:1708.09194 [cond-mat.dis-nn]}
  \BibitemShut {NoStop}%
\bibitem [{\citenamefont {Sethna}\ \emph {et~al.}(2017)\citenamefont {Sethna},
  \citenamefont {Bierbaum}, \citenamefont {Dahmen}, \citenamefont {Goodrich},
  \citenamefont {Greer}, \citenamefont {Hayden}, \citenamefont {Kent-Dobias},
  \citenamefont {Lee}, \citenamefont {Liarte}, \citenamefont {Ni},
  \citenamefont {Quinn}, \citenamefont {Raju}, \citenamefont {Rocklin},
  \citenamefont {Shekhawat},\ and\ \citenamefont {Zapperi}}]{sethna2017}%
  \BibitemOpen
  \bibfield  {author} {\bibinfo {author} {\bibfnamefont {J.~P.}\ \bibnamefont
  {Sethna}}, \bibinfo {author} {\bibfnamefont {M.~K.}\ \bibnamefont
  {Bierbaum}}, \bibinfo {author} {\bibfnamefont {K.~A.}\ \bibnamefont
  {Dahmen}}, \bibinfo {author} {\bibfnamefont {C.~P.}\ \bibnamefont
  {Goodrich}}, \bibinfo {author} {\bibfnamefont {J.~R.}\ \bibnamefont {Greer}},
  \bibinfo {author} {\bibfnamefont {L.~X.}\ \bibnamefont {Hayden}}, \bibinfo
  {author} {\bibfnamefont {J.~P.}\ \bibnamefont {Kent-Dobias}}, \bibinfo
  {author} {\bibfnamefont {E.~D.}\ \bibnamefont {Lee}}, \bibinfo {author}
  {\bibfnamefont {D.~B.}\ \bibnamefont {Liarte}}, \bibinfo {author}
  {\bibfnamefont {X.}~\bibnamefont {Ni}}, \bibinfo {author} {\bibfnamefont
  {K.~N.}\ \bibnamefont {Quinn}}, \bibinfo {author} {\bibfnamefont
  {A.}~\bibnamefont {Raju}}, \bibinfo {author} {\bibfnamefont {D.~Z.}\
  \bibnamefont {Rocklin}}, \bibinfo {author} {\bibfnamefont {A.}~\bibnamefont
  {Shekhawat}}, \ and\ \bibinfo {author} {\bibfnamefont {S.}~\bibnamefont
  {Zapperi}},\ }\href {\doibase 10.1146/annurev-matsci-070115-032036}
  {\bibfield  {journal} {\bibinfo  {journal} {Annual Review of Materials
  Research}\ }\textbf {\bibinfo {volume} {47}},\ \bibinfo {pages} {217}
  (\bibinfo {year} {2017})},\ \Eprint
  {http://arxiv.org/abs/https://doi.org/10.1146/annurev-matsci-070115-032036}
  {https://doi.org/10.1146/annurev-matsci-070115-032036} \BibitemShut {NoStop}%
\bibitem [{\citenamefont {Nattermann}\ and\ \citenamefont
  {Scheidl}(2000)}]{nattermann2000}%
  \BibitemOpen
  \bibfield  {author} {\bibinfo {author} {\bibfnamefont {T.}~\bibnamefont
  {Nattermann}}\ and\ \bibinfo {author} {\bibfnamefont {S.}~\bibnamefont
  {Scheidl}},\ }\href {\doibase 10.1080/000187300412257} {\bibfield  {journal}
  {\bibinfo  {journal} {Advances in Physics}\ }\textbf {\bibinfo {volume}
  {49}},\ \bibinfo {pages} {607} (\bibinfo {year} {2000})},\ \Eprint
  {http://arxiv.org/abs/https://doi.org/10.1080/000187300412257}
  {https://doi.org/10.1080/000187300412257} \BibitemShut {NoStop}%
\bibitem [{\citenamefont {{Giamarchi}}\ and\ \citenamefont
  {{Bhattacharya}}(2002)}]{giamarchi2002}%
  \BibitemOpen
  \bibfield  {author} {\bibinfo {author} {\bibfnamefont {T.}~\bibnamefont
  {{Giamarchi}}}\ and\ \bibinfo {author} {\bibfnamefont {S.}~\bibnamefont
  {{Bhattacharya}}},\ }in\ \href@noop {} {\emph {\bibinfo {booktitle} {High
  Magnetic Fields}}},\ \bibinfo {series} {Lecture Notes in Physics, Berlin
  Springer Verlag}, Vol.\ \bibinfo {volume} {595},\ \bibinfo {editor} {edited
  by\ \bibinfo {editor} {\bibfnamefont {C.}~\bibnamefont {{Berthier}}},
  \bibinfo {editor} {\bibfnamefont {L.~P.}\ \bibnamefont {{L{\'e}vy}}}, \ and\
  \bibinfo {editor} {\bibfnamefont {G.}~\bibnamefont {{Martinez}}}}\ (\bibinfo
  {year} {2002})\ pp.\ \bibinfo {pages} {314--360},\ \Eprint
  {http://arxiv.org/abs/cond-mat/0111052} {cond-mat/0111052} \BibitemShut
  {NoStop}%
\bibitem [{\citenamefont {Le~Doussal}(2010)}]{ledoussal2010}%
  \BibitemOpen
  \bibfield  {author} {\bibinfo {author} {\bibfnamefont {P.}~\bibnamefont
  {Le~Doussal}},\ }\href {\doibase 10.1142/S0217979210056384} {\bibfield
  {journal} {\bibinfo  {journal} {International Journal of Modern Physics B}\
  }\textbf {\bibinfo {volume} {24}},\ \bibinfo {pages} {3855} (\bibinfo {year}
  {2010})},\ \Eprint
  {http://arxiv.org/abs/https://www.worldscientific.com/doi/pdf/10.1142/S0217979210056384}
  {https://www.worldscientific.com/doi/pdf/10.1142/S0217979210056384}
  \BibitemShut {NoStop}%
\bibitem [{\citenamefont {Schulz}\ \emph {et~al.}(2012)\citenamefont {Schulz},
  \citenamefont {Ritz}, \citenamefont {Bauer}, \citenamefont {Halder},
  \citenamefont {Wagner}, \citenamefont {Franz}, \citenamefont {Pfleiderer},
  \citenamefont {Everschor}, \citenamefont {Garst},\ and\ \citenamefont
  {Rosch}}]{Schulz2012}%
  \BibitemOpen
  \bibfield  {author} {\bibinfo {author} {\bibfnamefont {T.}~\bibnamefont
  {Schulz}}, \bibinfo {author} {\bibfnamefont {R.}~\bibnamefont {Ritz}},
  \bibinfo {author} {\bibfnamefont {A.}~\bibnamefont {Bauer}}, \bibinfo
  {author} {\bibfnamefont {M.}~\bibnamefont {Halder}}, \bibinfo {author}
  {\bibfnamefont {M.}~\bibnamefont {Wagner}}, \bibinfo {author} {\bibfnamefont
  {C.}~\bibnamefont {Franz}}, \bibinfo {author} {\bibfnamefont
  {C.}~\bibnamefont {Pfleiderer}}, \bibinfo {author} {\bibfnamefont
  {K.}~\bibnamefont {Everschor}}, \bibinfo {author} {\bibfnamefont
  {M.}~\bibnamefont {Garst}}, \ and\ \bibinfo {author} {\bibfnamefont
  {A.}~\bibnamefont {Rosch}},\ }\href {http://dx.doi.org/10.1038/nphys2231}
  {\bibfield  {journal} {\bibinfo  {journal} {Nature Physics}\ }\textbf
  {\bibinfo {volume} {8}},\ \bibinfo {pages} {301 EP } (\bibinfo {year}
  {2012})}\BibitemShut {NoStop}%
\bibitem [{\citenamefont {Chepizhko}\ \emph {et~al.}(2016)\citenamefont
  {Chepizhko}, \citenamefont {Giampietro}, \citenamefont {Mastrapasqua},
  \citenamefont {Nourazar}, \citenamefont {Ascagni}, \citenamefont {Sugni},
  \citenamefont {Fascio}, \citenamefont {Leggio}, \citenamefont {Malinverno},
  \citenamefont {Scita}, \citenamefont {Santucci}, \citenamefont {Alava},
  \citenamefont {Zapperi},\ and\ \citenamefont {La~Porta}}]{Chepizhko2016}%
  \BibitemOpen
  \bibfield  {author} {\bibinfo {author} {\bibfnamefont {O.}~\bibnamefont
  {Chepizhko}}, \bibinfo {author} {\bibfnamefont {C.}~\bibnamefont
  {Giampietro}}, \bibinfo {author} {\bibfnamefont {E.}~\bibnamefont
  {Mastrapasqua}}, \bibinfo {author} {\bibfnamefont {M.}~\bibnamefont
  {Nourazar}}, \bibinfo {author} {\bibfnamefont {M.}~\bibnamefont {Ascagni}},
  \bibinfo {author} {\bibfnamefont {M.}~\bibnamefont {Sugni}}, \bibinfo
  {author} {\bibfnamefont {U.}~\bibnamefont {Fascio}}, \bibinfo {author}
  {\bibfnamefont {L.}~\bibnamefont {Leggio}}, \bibinfo {author} {\bibfnamefont
  {C.}~\bibnamefont {Malinverno}}, \bibinfo {author} {\bibfnamefont
  {G.}~\bibnamefont {Scita}}, \bibinfo {author} {\bibfnamefont
  {S.}~\bibnamefont {Santucci}}, \bibinfo {author} {\bibfnamefont {M.~J.}\
  \bibnamefont {Alava}}, \bibinfo {author} {\bibfnamefont {S.}~\bibnamefont
  {Zapperi}}, \ and\ \bibinfo {author} {\bibfnamefont {C.~A.~M.}\ \bibnamefont
  {La~Porta}},\ }\href {\doibase 10.1073/pnas.1600503113} {\bibfield  {journal}
  {\bibinfo  {journal} {Proceedings of the National Academy of Sciences}\
  }\textbf {\bibinfo {volume} {113}},\ \bibinfo {pages} {11408} (\bibinfo
  {year} {2016})},\ \Eprint
  {http://arxiv.org/abs/http://www.pnas.org/content/113/41/11408.full.pdf}
  {http://www.pnas.org/content/113/41/11408.full.pdf} \BibitemShut {NoStop}%
\bibitem [{\citenamefont {Jagla}\ and\ \citenamefont
  {Kolton}(2010)}]{jagla2010}%
  \BibitemOpen
  \bibfield  {author} {\bibinfo {author} {\bibfnamefont {E.~A.}\ \bibnamefont
  {Jagla}}\ and\ \bibinfo {author} {\bibfnamefont {A.~B.}\ \bibnamefont
  {Kolton}},\ }\href {\doibase 10.1029/2009JB006974} {\bibfield  {journal}
  {\bibinfo  {journal} {Journal of Geophysical Research: Solid Earth}\ }\textbf
  {\bibinfo {volume} {115}} (\bibinfo {year} {2010}),\ 10.1029/2009JB006974},\
  \Eprint
  {http://arxiv.org/abs/https://agupubs.onlinelibrary.wiley.com/doi/pdf/10.1029/2009JB006974}
  {https://agupubs.onlinelibrary.wiley.com/doi/pdf/10.1029/2009JB006974}
  \BibitemShut {NoStop}%
\bibitem [{\citenamefont {Jagla}\ \emph {et~al.}(2014)\citenamefont {Jagla},
  \citenamefont {Landes},\ and\ \citenamefont {Rosso}}]{jagla2014}%
  \BibitemOpen
  \bibfield  {author} {\bibinfo {author} {\bibfnamefont {E.~A.}\ \bibnamefont
  {Jagla}}, \bibinfo {author} {\bibfnamefont {F.~m. c.~P.}\ \bibnamefont
  {Landes}}, \ and\ \bibinfo {author} {\bibfnamefont {A.}~\bibnamefont
  {Rosso}},\ }\href {\doibase 10.1103/PhysRevLett.112.174301} {\bibfield
  {journal} {\bibinfo  {journal} {Phys. Rev. Lett.}\ }\textbf {\bibinfo
  {volume} {112}},\ \bibinfo {pages} {174301} (\bibinfo {year}
  {2014})}\BibitemShut {NoStop}%
\bibitem [{\citenamefont {{Fisher}}(1998)}]{Fisher1998}%
  \BibitemOpen
  \bibfield  {author} {\bibinfo {author} {\bibfnamefont {D.~S.}\ \bibnamefont
  {{Fisher}}},\ }\href {\doibase 10.1016/S0370-1573(98)00008-8} {\bibfield
  {journal} {\bibinfo  {journal} {Physics Reports}\ }\textbf {\bibinfo {volume}
  {301}},\ \bibinfo {pages} {113} (\bibinfo {year} {1998})},\ \Eprint
  {http://arxiv.org/abs/cond-mat/9711179} {cond-mat/9711179} \BibitemShut
  {NoStop}%
\bibitem [{\citenamefont {{Kardar}}(1998)}]{Kardar1998}%
  \BibitemOpen
  \bibfield  {author} {\bibinfo {author} {\bibfnamefont {M.}~\bibnamefont
  {{Kardar}}},\ }\href {\doibase 10.1016/S0370-1573(98)00007-6} {\bibfield
  {journal} {\bibinfo  {journal} {Physics Reports}\ }\textbf {\bibinfo {volume}
  {301}},\ \bibinfo {pages} {85} (\bibinfo {year} {1998})},\ \Eprint
  {http://arxiv.org/abs/cond-mat/9704172} {cond-mat/9704172} \BibitemShut
  {NoStop}%
\bibitem [{\citenamefont {Amaral}\ \emph {et~al.}(1994)\citenamefont {Amaral},
  \citenamefont {Barab\'asi},\ and\ \citenamefont {Stanley}}]{Amaral1994}%
  \BibitemOpen
  \bibfield  {author} {\bibinfo {author} {\bibfnamefont {L.~A.~N.}\
  \bibnamefont {Amaral}}, \bibinfo {author} {\bibfnamefont {A.-L.}\
  \bibnamefont {Barab\'asi}}, \ and\ \bibinfo {author} {\bibfnamefont {H.~E.}\
  \bibnamefont {Stanley}},\ }\href {\doibase 10.1103/PhysRevLett.73.62}
  {\bibfield  {journal} {\bibinfo  {journal} {Phys. Rev. Lett.}\ }\textbf
  {\bibinfo {volume} {73}},\ \bibinfo {pages} {62} (\bibinfo {year}
  {1994})}\BibitemShut {NoStop}%
\bibitem [{\citenamefont {Kolton}\ \emph
  {et~al.}(2006{\natexlab{a}})\citenamefont {Kolton}, \citenamefont {Rosso},
  \citenamefont {Giamarchi},\ and\ \citenamefont {Krauth}}]{kolton2006}%
  \BibitemOpen
  \bibfield  {author} {\bibinfo {author} {\bibfnamefont {A.~B.}\ \bibnamefont
  {Kolton}}, \bibinfo {author} {\bibfnamefont {A.}~\bibnamefont {Rosso}},
  \bibinfo {author} {\bibfnamefont {T.}~\bibnamefont {Giamarchi}}, \ and\
  \bibinfo {author} {\bibfnamefont {W.}~\bibnamefont {Krauth}},\ }\href
  {\doibase 10.1103/PhysRevLett.97.057001} {\bibfield  {journal} {\bibinfo
  {journal} {Phys. Rev. Lett.}\ }\textbf {\bibinfo {volume} {97}},\ \bibinfo
  {pages} {057001} (\bibinfo {year} {2006}{\natexlab{a}})}\BibitemShut
  {NoStop}%
\bibitem [{\citenamefont {Kolton}\ \emph
  {et~al.}(2009{\natexlab{a}})\citenamefont {Kolton}, \citenamefont {Rosso},
  \citenamefont {Giamarchi},\ and\ \citenamefont {Krauth}}]{kolton2009}%
  \BibitemOpen
  \bibfield  {author} {\bibinfo {author} {\bibfnamefont {A.~B.}\ \bibnamefont
  {Kolton}}, \bibinfo {author} {\bibfnamefont {A.}~\bibnamefont {Rosso}},
  \bibinfo {author} {\bibfnamefont {T.}~\bibnamefont {Giamarchi}}, \ and\
  \bibinfo {author} {\bibfnamefont {W.}~\bibnamefont {Krauth}},\ }\href
  {\doibase 10.1103/PhysRevB.79.184207} {\bibfield  {journal} {\bibinfo
  {journal} {Phys. Rev. B}\ }\textbf {\bibinfo {volume} {79}},\ \bibinfo
  {pages} {184207} (\bibinfo {year} {2009}{\natexlab{a}})}\BibitemShut
  {NoStop}%
\bibitem [{\citenamefont {Purrello}\ \emph {et~al.}(2017)\citenamefont
  {Purrello}, \citenamefont {Iguain}, \citenamefont {Kolton},\ and\
  \citenamefont {Jagla}}]{purrello2017}%
  \BibitemOpen
  \bibfield  {author} {\bibinfo {author} {\bibfnamefont {V.~H.}\ \bibnamefont
  {Purrello}}, \bibinfo {author} {\bibfnamefont {J.~L.}\ \bibnamefont
  {Iguain}}, \bibinfo {author} {\bibfnamefont {A.~B.}\ \bibnamefont {Kolton}},
  \ and\ \bibinfo {author} {\bibfnamefont {E.~A.}\ \bibnamefont {Jagla}},\
  }\href {\doibase 10.1103/PhysRevE.96.022112} {\bibfield  {journal} {\bibinfo
  {journal} {Phys. Rev. E}\ }\textbf {\bibinfo {volume} {96}},\ \bibinfo
  {pages} {022112} (\bibinfo {year} {2017})}\BibitemShut {NoStop}%
\bibitem [{\citenamefont {{Joerg Wiese}}\ and\ \citenamefont {{Le
  Doussal}}(2006)}]{wiese2006}%
  \BibitemOpen
  \bibfield  {author} {\bibinfo {author} {\bibfnamefont {K.}~\bibnamefont
  {{Joerg Wiese}}}\ and\ \bibinfo {author} {\bibfnamefont {P.}~\bibnamefont
  {{Le Doussal}}},\ }\href@noop {} {\bibfield  {journal} {\bibinfo  {journal}
  {eprint arXiv:cond-mat/0611346}\ } (\bibinfo {year} {2006})},\ \Eprint
  {http://arxiv.org/abs/cond-mat/0611346} {cond-mat/0611346} \BibitemShut
  {NoStop}%
\bibitem [{\citenamefont {Ferrero}\ \emph
  {et~al.}(2013{\natexlab{a}})\citenamefont {Ferrero}, \citenamefont
  {Bustingorry}, \citenamefont {Kolton},\ and\ \citenamefont
  {Rosso}}]{Ferrero2013}%
  \BibitemOpen
  \bibfield  {author} {\bibinfo {author} {\bibfnamefont {E.~E.}\ \bibnamefont
  {Ferrero}}, \bibinfo {author} {\bibfnamefont {S.}~\bibnamefont
  {Bustingorry}}, \bibinfo {author} {\bibfnamefont {A.~B.}\ \bibnamefont
  {Kolton}}, \ and\ \bibinfo {author} {\bibfnamefont {A.}~\bibnamefont
  {Rosso}},\ }\href {\doibase https://doi.org/10.1016/j.crhy.2013.08.002}
  {\bibfield  {journal} {\bibinfo  {journal} {Comptes Rendus Physique}\
  }\textbf {\bibinfo {volume} {14}},\ \bibinfo {pages} {641 } (\bibinfo {year}
  {2013}{\natexlab{a}})},\ \bibinfo {note} {disordered systems / Systèmes
  désordonnés}\BibitemShut {NoStop}%
\bibitem [{\citenamefont {Le~Doussal}\ \emph {et~al.}(2002)\citenamefont
  {Le~Doussal}, \citenamefont {Wiese},\ and\ \citenamefont
  {Chauve}}]{ledoussal2002}%
  \BibitemOpen
  \bibfield  {author} {\bibinfo {author} {\bibfnamefont {P.}~\bibnamefont
  {Le~Doussal}}, \bibinfo {author} {\bibfnamefont {K.~J.}\ \bibnamefont
  {Wiese}}, \ and\ \bibinfo {author} {\bibfnamefont {P.}~\bibnamefont
  {Chauve}},\ }\href {\doibase 10.1103/PhysRevB.66.174201} {\bibfield
  {journal} {\bibinfo  {journal} {Phys. Rev. B}\ }\textbf {\bibinfo {volume}
  {66}},\ \bibinfo {pages} {174201} (\bibinfo {year} {2002})}\BibitemShut
  {NoStop}%
\bibitem [{\citenamefont {Fedorenko}\ and\ \citenamefont
  {Stepanow}(2003)}]{Fedorenko2003}%
  \BibitemOpen
  \bibfield  {author} {\bibinfo {author} {\bibfnamefont {A.~A.}\ \bibnamefont
  {Fedorenko}}\ and\ \bibinfo {author} {\bibfnamefont {S.}~\bibnamefont
  {Stepanow}},\ }\href {\doibase 10.1103/PhysRevE.67.057104} {\bibfield
  {journal} {\bibinfo  {journal} {Phys. Rev. E}\ }\textbf {\bibinfo {volume}
  {67}},\ \bibinfo {pages} {057104} (\bibinfo {year} {2003})}\BibitemShut
  {NoStop}%
\bibitem [{\citenamefont {Leschhorn}(1993)}]{Leschhorn1993}%
  \BibitemOpen
  \bibfield  {author} {\bibinfo {author} {\bibfnamefont {H.}~\bibnamefont
  {Leschhorn}},\ }\href {\doibase https://doi.org/10.1016/0378-4371(93)90161-V}
  {\bibfield  {journal} {\bibinfo  {journal} {Physica A: Statistical Mechanics
  and its Applications}\ }\textbf {\bibinfo {volume} {195}},\ \bibinfo {pages}
  {324 } (\bibinfo {year} {1993})}\BibitemShut {NoStop}%
\bibitem [{\citenamefont {Leschhorn}\ \emph {et~al.}(1997)\citenamefont
  {Leschhorn}, \citenamefont {Nattermann}, \citenamefont {Stepanow},\ and\
  \citenamefont {Tang}}]{Leschhorn1997}%
  \BibitemOpen
  \bibfield  {author} {\bibinfo {author} {\bibfnamefont {H.}~\bibnamefont
  {Leschhorn}}, \bibinfo {author} {\bibfnamefont {T.}~\bibnamefont
  {Nattermann}}, \bibinfo {author} {\bibfnamefont {S.}~\bibnamefont
  {Stepanow}}, \ and\ \bibinfo {author} {\bibfnamefont {L.}~\bibnamefont
  {Tang}},\ }\href {\doibase 10.1002/andp.19975090102} {\bibfield  {journal}
  {\bibinfo  {journal} {Annalen der Physik}\ }\textbf {\bibinfo {volume}
  {509}},\ \bibinfo {pages} {1} (\bibinfo {year} {1997})},\ \Eprint
  {http://arxiv.org/abs/https://onlinelibrary.wiley.com/doi/pdf/10.1002/andp.19975090102}
  {https://onlinelibrary.wiley.com/doi/pdf/10.1002/andp.19975090102}
  \BibitemShut {NoStop}%
\bibitem [{\citenamefont {Roters}\ \emph {et~al.}(1999)\citenamefont {Roters},
  \citenamefont {Hucht}, \citenamefont {L\"ubeck}, \citenamefont {Nowak},\ and\
  \citenamefont {Usadel}}]{Roters1999}%
  \BibitemOpen
  \bibfield  {author} {\bibinfo {author} {\bibfnamefont {L.}~\bibnamefont
  {Roters}}, \bibinfo {author} {\bibfnamefont {A.}~\bibnamefont {Hucht}},
  \bibinfo {author} {\bibfnamefont {S.}~\bibnamefont {L\"ubeck}}, \bibinfo
  {author} {\bibfnamefont {U.}~\bibnamefont {Nowak}}, \ and\ \bibinfo {author}
  {\bibfnamefont {K.~D.}\ \bibnamefont {Usadel}},\ }\href {\doibase
  10.1103/PhysRevE.60.5202} {\bibfield  {journal} {\bibinfo  {journal} {Phys.
  Rev. E}\ }\textbf {\bibinfo {volume} {60}},\ \bibinfo {pages} {5202}
  (\bibinfo {year} {1999})}\BibitemShut {NoStop}%
\bibitem [{\citenamefont {Rosso}\ \emph {et~al.}(2003)\citenamefont {Rosso},
  \citenamefont {Hartmann},\ and\ \citenamefont {Krauth}}]{Rosso2003}%
  \BibitemOpen
  \bibfield  {author} {\bibinfo {author} {\bibfnamefont {A.}~\bibnamefont
  {Rosso}}, \bibinfo {author} {\bibfnamefont {A.~K.}\ \bibnamefont {Hartmann}},
  \ and\ \bibinfo {author} {\bibfnamefont {W.}~\bibnamefont {Krauth}},\ }\href
  {\doibase 10.1103/PhysRevE.67.021602} {\bibfield  {journal} {\bibinfo
  {journal} {Phys. Rev. E}\ }\textbf {\bibinfo {volume} {67}},\ \bibinfo
  {pages} {021602} (\bibinfo {year} {2003})}\BibitemShut {NoStop}%
\bibitem [{\citenamefont {Rosso}\ \emph {et~al.}(2007)\citenamefont {Rosso},
  \citenamefont {Le~Doussal},\ and\ \citenamefont {Wiese}}]{Rosso2007}%
  \BibitemOpen
  \bibfield  {author} {\bibinfo {author} {\bibfnamefont {A.}~\bibnamefont
  {Rosso}}, \bibinfo {author} {\bibfnamefont {P.}~\bibnamefont {Le~Doussal}}, \
  and\ \bibinfo {author} {\bibfnamefont {K.~J.}\ \bibnamefont {Wiese}},\ }\href
  {\doibase 10.1103/PhysRevB.75.220201} {\bibfield  {journal} {\bibinfo
  {journal} {Phys. Rev. B}\ }\textbf {\bibinfo {volume} {75}},\ \bibinfo
  {pages} {220201} (\bibinfo {year} {2007})}\BibitemShut {NoStop}%
\bibitem [{\citenamefont {Ferrero}\ \emph
  {et~al.}(2013{\natexlab{b}})\citenamefont {Ferrero}, \citenamefont
  {Bustingorry},\ and\ \citenamefont {Kolton}}]{Ferrero2013b}%
  \BibitemOpen
  \bibfield  {author} {\bibinfo {author} {\bibfnamefont {E.~E.}\ \bibnamefont
  {Ferrero}}, \bibinfo {author} {\bibfnamefont {S.}~\bibnamefont
  {Bustingorry}}, \ and\ \bibinfo {author} {\bibfnamefont {A.~B.}\ \bibnamefont
  {Kolton}},\ }\href {\doibase 10.1103/PhysRevE.87.032122} {\bibfield
  {journal} {\bibinfo  {journal} {Phys. Rev. E}\ }\textbf {\bibinfo {volume}
  {87}},\ \bibinfo {pages} {032122} (\bibinfo {year}
  {2013}{\natexlab{b}})}\BibitemShut {NoStop}%
\bibitem [{\citenamefont {Ramanathan}\ and\ \citenamefont
  {Fisher}(1998)}]{Ramanathan1998}%
  \BibitemOpen
  \bibfield  {author} {\bibinfo {author} {\bibfnamefont {S.}~\bibnamefont
  {Ramanathan}}\ and\ \bibinfo {author} {\bibfnamefont {D.~S.}\ \bibnamefont
  {Fisher}},\ }\href {\doibase 10.1103/PhysRevB.58.6026} {\bibfield  {journal}
  {\bibinfo  {journal} {Phys. Rev. B}\ }\textbf {\bibinfo {volume} {58}},\
  \bibinfo {pages} {6026} (\bibinfo {year} {1998})}\BibitemShut {NoStop}%
\bibitem [{\citenamefont {Zapperi}\ \emph {et~al.}(1998)\citenamefont
  {Zapperi}, \citenamefont {Cizeau}, \citenamefont {Durin},\ and\ \citenamefont
  {Stanley}}]{Zapperi1998}%
  \BibitemOpen
  \bibfield  {author} {\bibinfo {author} {\bibfnamefont {S.}~\bibnamefont
  {Zapperi}}, \bibinfo {author} {\bibfnamefont {P.}~\bibnamefont {Cizeau}},
  \bibinfo {author} {\bibfnamefont {G.}~\bibnamefont {Durin}}, \ and\ \bibinfo
  {author} {\bibfnamefont {H.~E.}\ \bibnamefont {Stanley}},\ }\href {\doibase
  10.1103/PhysRevB.58.6353} {\bibfield  {journal} {\bibinfo  {journal} {Phys.
  Rev. B}\ }\textbf {\bibinfo {volume} {58}},\ \bibinfo {pages} {6353}
  (\bibinfo {year} {1998})}\BibitemShut {NoStop}%
\bibitem [{\citenamefont {Rosso}\ and\ \citenamefont
  {Krauth}(2002)}]{Rosso2002}%
  \BibitemOpen
  \bibfield  {author} {\bibinfo {author} {\bibfnamefont {A.}~\bibnamefont
  {Rosso}}\ and\ \bibinfo {author} {\bibfnamefont {W.}~\bibnamefont {Krauth}},\
  }\href {\doibase 10.1103/PhysRevE.65.025101} {\bibfield  {journal} {\bibinfo
  {journal} {Phys. Rev. E}\ }\textbf {\bibinfo {volume} {65}},\ \bibinfo
  {pages} {025101} (\bibinfo {year} {2002})}\BibitemShut {NoStop}%
\bibitem [{\citenamefont {Duemmer}\ and\ \citenamefont
  {Krauth}(2007)}]{Duemmer2007}%
  \BibitemOpen
  \bibfield  {author} {\bibinfo {author} {\bibfnamefont {O.}~\bibnamefont
  {Duemmer}}\ and\ \bibinfo {author} {\bibfnamefont {W.}~\bibnamefont
  {Krauth}},\ }\href {http://stacks.iop.org/1742-5468/2007/i=01/a=P01019}
  {\bibfield  {journal} {\bibinfo  {journal} {Journal of Statistical Mechanics:
  Theory and Experiment}\ }\textbf {\bibinfo {volume} {2007}},\ \bibinfo
  {pages} {P01019} (\bibinfo {year} {2007})}\BibitemShut {NoStop}%
\bibitem [{\citenamefont {Laurson}\ \emph {et~al.}(2013)\citenamefont
  {Laurson}, \citenamefont {Illa}, \citenamefont {Santucci}, \citenamefont
  {Tore~Tallakstad}, \citenamefont {M{\aa}l{\o}y},\ and\ \citenamefont
  {Alava}}]{Laurson2013}%
  \BibitemOpen
  \bibfield  {author} {\bibinfo {author} {\bibfnamefont {L.}~\bibnamefont
  {Laurson}}, \bibinfo {author} {\bibfnamefont {X.}~\bibnamefont {Illa}},
  \bibinfo {author} {\bibfnamefont {S.}~\bibnamefont {Santucci}}, \bibinfo
  {author} {\bibfnamefont {K.}~\bibnamefont {Tore~Tallakstad}}, \bibinfo
  {author} {\bibfnamefont {K.~J.}\ \bibnamefont {M{\aa}l{\o}y}}, \ and\
  \bibinfo {author} {\bibfnamefont {M.~J.}\ \bibnamefont {Alava}},\ }\href
  {http://dx.doi.org/10.1038/ncomms3927} {\bibfield  {journal} {\bibinfo
  {journal} {Nature Communications}\ }\textbf {\bibinfo {volume} {4}},\
  \bibinfo {pages} {2927 EP } (\bibinfo {year} {2013})},\ \bibinfo {note}
  {article}\BibitemShut {NoStop}%
\bibitem [{\citenamefont {Boltz}\ and\ \citenamefont
  {Kierfeld}(2014)}]{Boltz2014}%
  \BibitemOpen
  \bibfield  {author} {\bibinfo {author} {\bibfnamefont {H.-H.}\ \bibnamefont
  {Boltz}}\ and\ \bibinfo {author} {\bibfnamefont {J.}~\bibnamefont
  {Kierfeld}},\ }\href {\doibase 10.1103/PhysRevE.90.012101} {\bibfield
  {journal} {\bibinfo  {journal} {Phys. Rev. E}\ }\textbf {\bibinfo {volume}
  {90}},\ \bibinfo {pages} {012101} (\bibinfo {year} {2014})}\BibitemShut
  {NoStop}%
\bibitem [{\citenamefont {Tang}\ \emph {et~al.}(1995)\citenamefont {Tang},
  \citenamefont {Kardar},\ and\ \citenamefont {Dhar}}]{Tang1995}%
  \BibitemOpen
  \bibfield  {author} {\bibinfo {author} {\bibfnamefont {L.-H.}\ \bibnamefont
  {Tang}}, \bibinfo {author} {\bibfnamefont {M.}~\bibnamefont {Kardar}}, \ and\
  \bibinfo {author} {\bibfnamefont {D.}~\bibnamefont {Dhar}},\ }\href {\doibase
  10.1103/PhysRevLett.74.920} {\bibfield  {journal} {\bibinfo  {journal} {Phys.
  Rev. Lett.}\ }\textbf {\bibinfo {volume} {74}},\ \bibinfo {pages} {920}
  (\bibinfo {year} {1995})}\BibitemShut {NoStop}%
\bibitem [{\citenamefont {Fedorenko}\ \emph {et~al.}(2006)\citenamefont
  {Fedorenko}, \citenamefont {Le~Doussal},\ and\ \citenamefont
  {Wiese}}]{Fedorenko2006}%
  \BibitemOpen
  \bibfield  {author} {\bibinfo {author} {\bibfnamefont {A.~A.}\ \bibnamefont
  {Fedorenko}}, \bibinfo {author} {\bibfnamefont {P.}~\bibnamefont
  {Le~Doussal}}, \ and\ \bibinfo {author} {\bibfnamefont {K.~J.}\ \bibnamefont
  {Wiese}},\ }\href {\doibase 10.1103/PhysRevE.74.061109} {\bibfield  {journal}
  {\bibinfo  {journal} {Phys. Rev. E}\ }\textbf {\bibinfo {volume} {74}},\
  \bibinfo {pages} {061109} (\bibinfo {year} {2006})}\BibitemShut {NoStop}%
\bibitem [{\citenamefont {Bustingorry}\ \emph {et~al.}(2010)\citenamefont
  {Bustingorry}, \citenamefont {Kolton},\ and\ \citenamefont
  {Giamarchi}}]{Bustingorry2010}%
  \BibitemOpen
  \bibfield  {author} {\bibinfo {author} {\bibfnamefont {S.}~\bibnamefont
  {Bustingorry}}, \bibinfo {author} {\bibfnamefont {A.~B.}\ \bibnamefont
  {Kolton}}, \ and\ \bibinfo {author} {\bibfnamefont {T.}~\bibnamefont
  {Giamarchi}},\ }\href {\doibase 10.1103/PhysRevB.82.094202} {\bibfield
  {journal} {\bibinfo  {journal} {Phys. Rev. B}\ }\textbf {\bibinfo {volume}
  {82}},\ \bibinfo {pages} {094202} (\bibinfo {year} {2010})}\BibitemShut
  {NoStop}%
\bibitem [{\citenamefont {Rosso}\ and\ \citenamefont
  {Krauth}(2001{\natexlab{a}})}]{Rosso2001b}%
  \BibitemOpen
  \bibfield  {author} {\bibinfo {author} {\bibfnamefont {A.}~\bibnamefont
  {Rosso}}\ and\ \bibinfo {author} {\bibfnamefont {W.}~\bibnamefont {Krauth}},\
  }\href {\doibase 10.1103/PhysRevLett.87.187002} {\bibfield  {journal}
  {\bibinfo  {journal} {Phys. Rev. Lett.}\ }\textbf {\bibinfo {volume} {87}},\
  \bibinfo {pages} {187002} (\bibinfo {year} {2001}{\natexlab{a}})}\BibitemShut
  {NoStop}%
\bibitem [{\citenamefont {Goodman}\ and\ \citenamefont
  {Teitel}(2004)}]{Goodman2004}%
  \BibitemOpen
  \bibfield  {author} {\bibinfo {author} {\bibfnamefont {T.}~\bibnamefont
  {Goodman}}\ and\ \bibinfo {author} {\bibfnamefont {S.}~\bibnamefont
  {Teitel}},\ }\href {\doibase 10.1103/PhysRevE.69.062105} {\bibfield
  {journal} {\bibinfo  {journal} {Phys. Rev. E}\ }\textbf {\bibinfo {volume}
  {69}},\ \bibinfo {pages} {062105} (\bibinfo {year} {2004})}\BibitemShut
  {NoStop}%
\bibitem [{\citenamefont {Le~Doussal}\ and\ \citenamefont
  {Wiese}(2003)}]{ledoussal2003}%
  \BibitemOpen
  \bibfield  {author} {\bibinfo {author} {\bibfnamefont {P.}~\bibnamefont
  {Le~Doussal}}\ and\ \bibinfo {author} {\bibfnamefont {K.~J.}\ \bibnamefont
  {Wiese}},\ }\href {\doibase 10.1103/PhysRevE.67.016121} {\bibfield  {journal}
  {\bibinfo  {journal} {Phys. Rev. E}\ }\textbf {\bibinfo {volume} {67}},\
  \bibinfo {pages} {016121} (\bibinfo {year} {2003})}\BibitemShut {NoStop}%
\bibitem [{\citenamefont {Chen}\ \emph {et~al.}(2015)\citenamefont {Chen},
  \citenamefont {Zapperi},\ and\ \citenamefont {Sethna}}]{Chen2015}%
  \BibitemOpen
  \bibfield  {author} {\bibinfo {author} {\bibfnamefont {Y.~J.}\ \bibnamefont
  {Chen}}, \bibinfo {author} {\bibfnamefont {S.}~\bibnamefont {Zapperi}}, \
  and\ \bibinfo {author} {\bibfnamefont {J.~P.}\ \bibnamefont {Sethna}},\
  }\href {\doibase 10.1103/PhysRevE.92.022146} {\bibfield  {journal} {\bibinfo
  {journal} {Phys. Rev. E}\ }\textbf {\bibinfo {volume} {92}},\ \bibinfo
  {pages} {022146} (\bibinfo {year} {2015})}\BibitemShut {NoStop}%
\bibitem [{\citenamefont {Arag\'on}\ \emph {et~al.}(2016)\citenamefont
  {Arag\'on}, \citenamefont {Kolton}, \citenamefont {Doussal}, \citenamefont
  {Wiese},\ and\ \citenamefont {Jagla}}]{Aragon2016}%
  \BibitemOpen
  \bibfield  {author} {\bibinfo {author} {\bibfnamefont {L.~E.}\ \bibnamefont
  {Arag\'on}}, \bibinfo {author} {\bibfnamefont {A.~B.}\ \bibnamefont
  {Kolton}}, \bibinfo {author} {\bibfnamefont {P.~L.}\ \bibnamefont {Doussal}},
  \bibinfo {author} {\bibfnamefont {K.~J.}\ \bibnamefont {Wiese}}, \ and\
  \bibinfo {author} {\bibfnamefont {E.~A.}\ \bibnamefont {Jagla}},\ }\href
  {http://stacks.iop.org/0295-5075/113/i=1/a=10002} {\bibfield  {journal}
  {\bibinfo  {journal} {EPL (Europhysics Letters)}\ }\textbf {\bibinfo {volume}
  {113}},\ \bibinfo {pages} {10002} (\bibinfo {year} {2016})}\BibitemShut
  {NoStop}%
\bibitem [{\citenamefont {Glatz}\ \emph {et~al.}(2003)\citenamefont {Glatz},
  \citenamefont {Nattermann},\ and\ \citenamefont {Pokrovsky}}]{glatz2003}%
  \BibitemOpen
  \bibfield  {author} {\bibinfo {author} {\bibfnamefont {A.}~\bibnamefont
  {Glatz}}, \bibinfo {author} {\bibfnamefont {T.}~\bibnamefont {Nattermann}}, \
  and\ \bibinfo {author} {\bibfnamefont {V.}~\bibnamefont {Pokrovsky}},\ }\href
  {\doibase 10.1103/PhysRevLett.90.047201} {\bibfield  {journal} {\bibinfo
  {journal} {Phys. Rev. Lett.}\ }\textbf {\bibinfo {volume} {90}},\ \bibinfo
  {pages} {047201} (\bibinfo {year} {2003})}\BibitemShut {NoStop}%
\bibitem [{Note1()}]{Note1}%
  \BibitemOpen
  \bibinfo {note} {We already discard the possible existence of finite-size
  ~\cite {Duemmer2005} and other type of crossovers (see e.g. Refs.~\cite
  {Bustingorry2010,Bustingorry2010b,Chen2015}), or thermal rounding
  effects~\cite {Chen1995,Bustingorry2012,purrello2017} which may also
  difficult the experimental observation of a clear power-law in the velocity
  force characteristics.}\BibitemShut {Stop}%
\bibitem [{\citenamefont {Bolech}\ and\ \citenamefont
  {Rosso}(2004)}]{Bolech2004}%
  \BibitemOpen
  \bibfield  {author} {\bibinfo {author} {\bibfnamefont {C.~J.}\ \bibnamefont
  {Bolech}}\ and\ \bibinfo {author} {\bibfnamefont {A.}~\bibnamefont {Rosso}},\
  }\href {\doibase 10.1103/PhysRevLett.93.125701} {\bibfield  {journal}
  {\bibinfo  {journal} {Phys. Rev. Lett.}\ }\textbf {\bibinfo {volume} {93}},\
  \bibinfo {pages} {125701} (\bibinfo {year} {2004})}\BibitemShut {NoStop}%
\bibitem [{\citenamefont {Kolton}\ \emph {et~al.}(2013)\citenamefont {Kolton},
  \citenamefont {Bustingorry}, \citenamefont {Ferrero},\ and\ \citenamefont
  {Rosso}}]{Kolton2013}%
  \BibitemOpen
  \bibfield  {author} {\bibinfo {author} {\bibfnamefont {A.~B.}\ \bibnamefont
  {Kolton}}, \bibinfo {author} {\bibfnamefont {S.}~\bibnamefont {Bustingorry}},
  \bibinfo {author} {\bibfnamefont {E.~E.}\ \bibnamefont {Ferrero}}, \ and\
  \bibinfo {author} {\bibfnamefont {A.}~\bibnamefont {Rosso}},\ }\href
  {http://stacks.iop.org/1742-5468/2013/i=12/a=P12004} {\bibfield  {journal}
  {\bibinfo  {journal} {Journal of Statistical Mechanics: Theory and
  Experiment}\ }\textbf {\bibinfo {volume} {2013}},\ \bibinfo {pages} {P12004}
  (\bibinfo {year} {2013})}\BibitemShut {NoStop}%
\bibitem [{Note2()}]{Note2}%
  \BibitemOpen
  \bibinfo {note} {It is interesting in this respect to quote the following
  statements by Fisher, in Ref.\protect \rev@citealp {Fisher1998}: ``The
  velocity exponent, $\beta $, in mean field theory is not fully universal. It
  depends on whether or not the pinning forces are continuously differentiable
  and, if not, on the nature of the singularities in $f_p(u)$. For all but
  discontinuous forces, this causes long time scales in mean-field theory
  associated with the acceleration away from configurations that have just
  become unstable. With short range interactions, these should not play a role
  due to the jerky nature of $\phi ({\protect \bf r},t)$ caused by jumps. Thus
  the mean-field models with $\beta _{MF}=1$ are the \protect \textit {right}
  ones about which to attempt an RG analysis.''}\BibitemShut {NoStop}%
\bibitem [{Note3()}]{Note3}%
  \BibitemOpen
  \bibinfo {note} {We assume that $f_c$ is not controlled by extreme statistics
  in the large-size limit~\cite {Kolton2013}.}\BibitemShut {Stop}%
\bibitem [{\citenamefont {Middleton}(1992)}]{Middleton1992}%
  \BibitemOpen
  \bibfield  {author} {\bibinfo {author} {\bibfnamefont {A.~A.}\ \bibnamefont
  {Middleton}},\ }\href {\doibase 10.1103/PhysRevLett.68.670} {\bibfield
  {journal} {\bibinfo  {journal} {Phys. Rev. Lett.}\ }\textbf {\bibinfo
  {volume} {68}},\ \bibinfo {pages} {670} (\bibinfo {year} {1992})}\BibitemShut
  {NoStop}%
\bibitem [{\citenamefont {Rosso}\ and\ \citenamefont
  {Krauth}(2001{\natexlab{b}})}]{Rosso2001}%
  \BibitemOpen
  \bibfield  {author} {\bibinfo {author} {\bibfnamefont {A.}~\bibnamefont
  {Rosso}}\ and\ \bibinfo {author} {\bibfnamefont {W.}~\bibnamefont {Krauth}},\
  }\href {\doibase 10.1103/PhysRevB.65.012202} {\bibfield  {journal} {\bibinfo
  {journal} {Phys. Rev. B}\ }\textbf {\bibinfo {volume} {65}},\ \bibinfo
  {pages} {012202} (\bibinfo {year} {2001}{\natexlab{b}})}\BibitemShut
  {NoStop}%
\bibitem [{\citenamefont {Rosso}\ and\ \citenamefont
  {Krauth}(2005)}]{Rosso2005}%
  \BibitemOpen
  \bibfield  {author} {\bibinfo {author} {\bibfnamefont {A.}~\bibnamefont
  {Rosso}}\ and\ \bibinfo {author} {\bibfnamefont {W.}~\bibnamefont {Krauth}},\
  }\href {\doibase https://doi.org/10.1016/j.cpc.2005.03.042} {\bibfield
  {journal} {\bibinfo  {journal} {Computer Physics Communications}\ }\textbf
  {\bibinfo {volume} {169}},\ \bibinfo {pages} {188 } (\bibinfo {year}
  {2005})},\ \bibinfo {note} {proceedings of the Europhysics Conference on
  Computational Physics 2004}\BibitemShut {NoStop}%
\bibitem [{\citenamefont {Kolton}\ \emph
  {et~al.}(2009{\natexlab{b}})\citenamefont {Kolton}, \citenamefont {Schehr},\
  and\ \citenamefont {Le~Doussal}}]{Kolton2009b}%
  \BibitemOpen
  \bibfield  {author} {\bibinfo {author} {\bibfnamefont {A.~B.}\ \bibnamefont
  {Kolton}}, \bibinfo {author} {\bibfnamefont {G.}~\bibnamefont {Schehr}}, \
  and\ \bibinfo {author} {\bibfnamefont {P.}~\bibnamefont {Le~Doussal}},\
  }\href {\doibase 10.1103/PhysRevLett.103.160602} {\bibfield  {journal}
  {\bibinfo  {journal} {Phys. Rev. Lett.}\ }\textbf {\bibinfo {volume} {103}},\
  \bibinfo {pages} {160602} (\bibinfo {year} {2009}{\natexlab{b}})}\BibitemShut
  {NoStop}%
\bibitem [{Note4()}]{Note4}%
  \BibitemOpen
  \bibinfo {note} {The discussion of anisotropic depinning universality
  classes, where STS is broken will be published elsewhere. Nevertheless, we
  believe that our general conclusions hold also for this case.}\BibitemShut
  {Stop}%
\bibitem [{\citenamefont {Kolton}\ \emph
  {et~al.}(2006{\natexlab{b}})\citenamefont {Kolton}, \citenamefont {Rosso},
  \citenamefont {Albano},\ and\ \citenamefont {Giamarchi}}]{kolton2006b}%
  \BibitemOpen
  \bibfield  {author} {\bibinfo {author} {\bibfnamefont {A.~B.}\ \bibnamefont
  {Kolton}}, \bibinfo {author} {\bibfnamefont {A.}~\bibnamefont {Rosso}},
  \bibinfo {author} {\bibfnamefont {E.~V.}\ \bibnamefont {Albano}}, \ and\
  \bibinfo {author} {\bibfnamefont {T.}~\bibnamefont {Giamarchi}},\ }\href
  {\doibase 10.1103/PhysRevB.74.140201} {\bibfield  {journal} {\bibinfo
  {journal} {Phys. Rev. B}\ }\textbf {\bibinfo {volume} {74}},\ \bibinfo
  {pages} {140201} (\bibinfo {year} {2006}{\natexlab{b}})}\BibitemShut
  {NoStop}%
\bibitem [{\citenamefont {Chauve}\ \emph {et~al.}(2000)\citenamefont {Chauve},
  \citenamefont {Giamarchi},\ and\ \citenamefont {Le~Doussal}}]{chauve2000}%
  \BibitemOpen
  \bibfield  {author} {\bibinfo {author} {\bibfnamefont {P.}~\bibnamefont
  {Chauve}}, \bibinfo {author} {\bibfnamefont {T.}~\bibnamefont {Giamarchi}}, \
  and\ \bibinfo {author} {\bibfnamefont {P.}~\bibnamefont {Le~Doussal}},\
  }\href {\doibase 10.1103/PhysRevB.62.6241} {\bibfield  {journal} {\bibinfo
  {journal} {Phys. Rev. B}\ }\textbf {\bibinfo {volume} {62}},\ \bibinfo
  {pages} {6241} (\bibinfo {year} {2000})}\BibitemShut {NoStop}%
\bibitem [{Note5()}]{Note5}%
  \BibitemOpen
  \bibinfo {note} {The $\sigma =d/2$ case has $\beta =1$ but with weak
  logarithmic corrections ~\cite {Fedorenko2003}.}\BibitemShut {Stop}%
\bibitem [{\citenamefont {Popov}\ and\ \citenamefont {Gray}(2014)}]{Popov2014}%
  \BibitemOpen
  \bibfield  {author} {\bibinfo {author} {\bibfnamefont {V.~L.}\ \bibnamefont
  {Popov}}\ and\ \bibinfo {author} {\bibfnamefont {J.~A.~T.}\ \bibnamefont
  {Gray}},\ }in\ \href {\doibase 10.1007/978-3-642-39905-3_10} {\emph {\bibinfo
  {booktitle} {The History of Theoretical, Material and Computational Mechanics
  - Mathematics Meets Mechanics and Engineering}}}\ (\bibinfo  {publisher}
  {Springer Berlin Heidelberg},\ \bibinfo {year} {2014})\ pp.\ \bibinfo {pages}
  {153--168}\BibitemShut {NoStop}%
\bibitem [{\citenamefont {Cao}\ \emph {et~al.}(2018)\citenamefont {Cao},
  \citenamefont {Bouzat}, \citenamefont {Kolton},\ and\ \citenamefont
  {Rosso}}]{Cao2018}%
  \BibitemOpen
  \bibfield  {author} {\bibinfo {author} {\bibfnamefont {X.}~\bibnamefont
  {Cao}}, \bibinfo {author} {\bibfnamefont {S.}~\bibnamefont {Bouzat}},
  \bibinfo {author} {\bibfnamefont {A.~B.}\ \bibnamefont {Kolton}}, \ and\
  \bibinfo {author} {\bibfnamefont {A.}~\bibnamefont {Rosso}},\ }\href
  {\doibase 10.1103/PhysRevE.97.022118} {\bibfield  {journal} {\bibinfo
  {journal} {Phys. Rev. E}\ }\textbf {\bibinfo {volume} {97}},\ \bibinfo
  {pages} {022118} (\bibinfo {year} {2018})}\BibitemShut {NoStop}%
\bibitem [{Note6()}]{Note6}%
  \BibitemOpen
  \bibinfo {note} {The value for $\sigma =2$, is somewhat lower than the
  expected $\beta =1$, however this is not surprising in view of the
  logarithmic corrections expected in this case (see \cite
  {Fedorenko2003}).}\BibitemShut {Stop}%
\bibitem [{\citenamefont {{Fern{\'a}ndez Aguirre}}\ and\ \citenamefont
  {{Jagla}}(2018)}]{Aguirre2018}%
  \BibitemOpen
  \bibfield  {author} {\bibinfo {author} {\bibfnamefont {I.}~\bibnamefont
  {{Fern{\'a}ndez Aguirre}}}\ and\ \bibinfo {author} {\bibfnamefont {E.~A.}\
  \bibnamefont {{Jagla}}},\ }\href@noop {} {\bibfield  {journal} {\bibinfo
  {journal} {ArXiv e-prints}\ } (\bibinfo {year} {2018})},\ \Eprint
  {http://arxiv.org/abs/1803.10072} {arXiv:1803.10072 [cond-mat.stat-mech]}
  \BibitemShut {NoStop}%
\bibitem [{Note7()}]{Note7}%
  \BibitemOpen
  \bibinfo {note} {I. Fern\'andez Aguirre, A. Rosso, and E. A. Jagla,
  unpublished.}\BibitemShut {Stop}%
\bibitem [{\citenamefont {Duemmer}\ and\ \citenamefont
  {Krauth}(2005)}]{Duemmer2005}%
  \BibitemOpen
  \bibfield  {author} {\bibinfo {author} {\bibfnamefont {O.}~\bibnamefont
  {Duemmer}}\ and\ \bibinfo {author} {\bibfnamefont {W.}~\bibnamefont
  {Krauth}},\ }\href {\doibase 10.1103/PhysRevE.71.061601} {\bibfield
  {journal} {\bibinfo  {journal} {Phys. Rev. E}\ }\textbf {\bibinfo {volume}
  {71}},\ \bibinfo {pages} {061601} (\bibinfo {year} {2005})}\BibitemShut
  {NoStop}%
\bibitem [{\citenamefont {Bustingorry}\ and\ \citenamefont
  {Kolton}(2010)}]{Bustingorry2010b}%
  \BibitemOpen
  \bibfield  {author} {\bibinfo {author} {\bibfnamefont {S.}~\bibnamefont
  {Bustingorry}}\ and\ \bibinfo {author} {\bibfnamefont {A.}~\bibnamefont
  {Kolton}},\ }\href
  {http://www.papersinphysics.org/papersinphysics/article/view/44} {\bibfield
  {journal} {\bibinfo  {journal} {Papers in Physics}\ }\textbf {\bibinfo
  {volume} {2}} (\bibinfo {year} {2010})}\BibitemShut {NoStop}%
\bibitem [{\citenamefont {Chen}\ and\ \citenamefont
  {Marchetti}(1995)}]{Chen1995}%
  \BibitemOpen
  \bibfield  {author} {\bibinfo {author} {\bibfnamefont {L.-W.}\ \bibnamefont
  {Chen}}\ and\ \bibinfo {author} {\bibfnamefont {M.~C.}\ \bibnamefont
  {Marchetti}},\ }\href {\doibase 10.1103/PhysRevB.51.6296} {\bibfield
  {journal} {\bibinfo  {journal} {Phys. Rev. B}\ }\textbf {\bibinfo {volume}
  {51}},\ \bibinfo {pages} {6296} (\bibinfo {year} {1995})}\BibitemShut
  {NoStop}%
\bibitem [{\citenamefont {Bustingorry}\ \emph {et~al.}(2012)\citenamefont
  {Bustingorry}, \citenamefont {Kolton},\ and\ \citenamefont
  {Giamarchi}}]{Bustingorry2012}%
  \BibitemOpen
  \bibfield  {author} {\bibinfo {author} {\bibfnamefont {S.}~\bibnamefont
  {Bustingorry}}, \bibinfo {author} {\bibfnamefont {A.~B.}\ \bibnamefont
  {Kolton}}, \ and\ \bibinfo {author} {\bibfnamefont {T.}~\bibnamefont
  {Giamarchi}},\ }\href {\doibase 10.1103/PhysRevE.85.021144} {\bibfield
  {journal} {\bibinfo  {journal} {Phys. Rev. E}\ }\textbf {\bibinfo {volume}
  {85}},\ \bibinfo {pages} {021144} (\bibinfo {year} {2012})}\BibitemShut
  {NoStop}%
\end{thebibliography}%

\tableofcontents
\end{document}